\documentclass[aps,pre,twocolumn,unsortedaddress,floatfix]{revtex4}
\usepackage[utf8]{inputenc}
\usepackage{graphicx}
\usepackage{amsmath}
\usepackage{amssymb}
\usepackage{comment}
\usepackage[dvipsnames]{xcolor}
\usepackage{chngcntr}

\usepackage[caption=false,position=top]{subfig}

\makeatletter
\def\maketitle{
\@author@finish
\title@column\titleblock@produce
\suppressfloats[t]}
\makeatother

\begin{document}

\title{Grain boundary segregation and phase separation in ceria-zirconia from atomistic simulation}
\author{Tom L. Underwood$^{1,2}$}
\author{Susanna Vigorito$^3$}
\author{Marco Molinari$^3$}
\author{John Purton$^4$}
\author{Nigel B. Wilding$^5$}
\author{John T. S. Irvine$^6$}
\author{Stephen C. Parker$^1$}
\affiliation{$^1$Department of Chemistry, University of Bath, Claverton Down, Bath, BA2 7AY. U.K.}
\affiliation{$^2$Scientific Computing Department STFC, Rutherford Appleton Laboratory, Harwell Campus, Didcot, OX11 0QX, U.K.}
\affiliation{$^3$Department of Chemical Sciences, University of Huddersfield, Queensgate, Huddersfield, HD1 3DH, U.K.}
\affiliation{$^4$Scientiﬁc Computing Department, STFC, Daresbury Laboratory, Keckwick Lane, Warrington, WA4 4AD, U.K.}
\affiliation{$^5$H.H. Wills Physics Laboratory, University of Bristol, Royal Fort, Bristol, BS8 1TL, U.K.}
\affiliation{$^6$School of Chemistry, University of St Andrews, St Andrews, KY16 9ST, U.K.}

\begin{abstract}
Doping is the most common strategy employed in the development of new and improved materials. However, predicting the effects of doping on the 
atomic-scale structure of a material is often difficult or limited to high-end experimental techniques. 
Doping can induce phase separation in a material, undermining the material's stability. A further complication is that dopant atoms can segregate to 
interfaces in a material such as grain boundaries (GBs), with consequences for key macroscopic properties of the material such as its conductivity.
Here, we describe a computational methodology based on semi-grand canonical Monte Carlo which can be used to probe these phenomena at the
atomic scale for metal oxide solid solutions. The methodology can provide precise predictions of the thermodynamic conditions at which phase separation 
occurs. It can also provide the segregation patterns exhibited by GBs at given conditions. We apply the methodology to one of the most important 
catalytic materials, ceria-zirconia. Our calculations reveal an interesting richness in the GB segregation in this system. Most GBs we examined exhibited 
continuous increases in Zr segregation upon Zr doping, with a concomitant reduction in the formation enthalpies of the GBs. However, a few GBs exhibited 
no segregation at low temperatures. We also observed evidence of first-order complexion transitions in some GBs.
\end{abstract}

\maketitle

\section{Introduction}\label{sec:intro}
It is well known that the atomic-scale structure of grain boundaries (GBs) in a material can strongly influence
macroscopic properties of the material such as strength and conductivity \cite{Mittemeijer2021,Lejcek2010,Priester2013}. 
For this reason, there is considerable interest in being able to control GB structure at the atomic scale
in order to design new materials with superior properties \cite{Harmer2011,Raabe2014,Watanabe2011}. 
One way in which GB structure can be controlled is by adding dopants to a material \cite{Raabe2014,Watanabe2011}. 
Depending on the temperature, dopant concentration, and the particular GB type, dopant atoms may \emph{segregate} to GBs, 
bringing about changes in their structure and, ultimately, the macroscopic properties of the material \cite{Lejcek2010,Lejcek2017}. 
However, this phenomenon is nontrivial, for which reason there is interest in using computer simulation to obtain accurate maps of 
GB properties over a range of thermodynamic parameters \cite{Shi2011,Wang2021}.

Analytical and field-based \cite{Lejcek2010,Priester2013,Darvishi2020,Wang2021} models have added great insight into 
GB properties including segregation in solid solutions. However, they cannot provide a detailed description of GB structure at the atomic scale. 
On the other hand, density-functional theory (DFT) \cite{Hohenberg1964,Martin2020} is in principle capable of providing such detail with 
high accuracy \cite{Grau-Crespo2011}. However, DFT is in practice limited to small system sizes on account of its computational expense.
A middle ground is found in using interatomic potentials, which strike a balance between accuracy and computational cost. 
Using interatomic potentials, molecular dynamics (MD)\cite{Allen1989} and Monte Carlo (MC)\cite{Frenkel2002,Dubbeldam2013} simulation have
been used to study GBs in solid solutions. While MD is a powerful method for quantifying time-dependent 
properties such as defect diffusion coefficients, in solid solutions it can prove intractable with MD to reach GB structures corresponding to 
thermodynamic equilibrium due to long timescales associated with dopant diffusion. 
MC can sidestep this issue by utilising \emph{unphysical} dynamics which enable equilibrium to be reached quickly. 
For instance, to study segregation in metal oxide solid solutions, dynamics which involve swapping the positions of atoms belonging to different 
elements has been successfully utilised \cite{Yoshiya2011,Lee2013,Lee2010,Purton2019,Gunn2018,Purton2017,Purton2006,Purton2005}.
Moreover, the semi-grand canonical MC (SGCMC)\cite{Kofke1988,Frenkel2002} method,
which utilises dynamics where the elements of atoms are \emph{transformed} in-place during the simulation, has proved to be a powerful probe of
equilibrium GB structures in alloys
\cite{Frolov2013,Frolov2015,Pan2016,Yang2018,Koju2020PRM,Koju2020AM,Sansoz2022,Hu2019,Hu2020,Hu2021}. 

While, as mentioned above, the SGCMC method has been used to study GBs in alloys, 
as far as we are aware it has yet to be applied to GBs in \emph{metal oxide solid solutions}. 
Given the importance of this class of material to technology, and the success SGCMC has had in providing insight into alloy GBs, 
this is something we believe deserves attention.
Moreover, while the use of SGCMC to study GBs is a relatively recent development, SGCMC has a long history of being used to calculate phase diagrams
of mixtures, a task for which it is particularly well suited \cite{Frenkel2002}. 
SGCMC has even been used in conjunction with \emph{free energy methods}\cite{Bruce2003,Iba2001} -- i.e. methods which utilise adaptive algorithms to learn
free energy landscapes in the vicinity of phase transitions -- in order to calculate phase diagrams to high precision \cite{Shen2006,Wilding2010}.
An interesting prospect is to adapt these methods to enable high-precision calculations of phase diagrams of metal oxide solid solutions.

SGCMC thus has the potential to be a powerful tool for investigating GB structure, and, in conjunction with free energy methods, calculating precise bulk 
phase diagrams, in metal oxide solid solutions. The aim of this work is to take steps towards realising this. As a testing ground we consider the
material ceria-zirconia, Ce$_{1-c}$Zr$_c$O$_2$.
Ceria-zirconia has myriad catalytic applications, most notably as an oxygen buffer in three-way catalytic converters on account of its oxygen storage 
capacity \cite{Devaiah2018}.
There hence is considerable interest in understanding its phase diagram and atomic-scale structure. 
Accordingly its bulk phase diagram has been the subject of numerous experimental 
\cite{Duwez1950,Tani1983,Duran1990,Bozo2001,Cabanas2001,Dutta2006,Kim2006,Lee2008,Oh2022JPE} 
and computational studies \cite{Balducci1997,Balducci1998,Bozo2001,Conesa2003,Dutta2006,Grau-Crespo2011,Oh2022AEM}. 
Studies suggest that for a wide range of temperatures and compositions ceria-zirconia is thermodynamically unstable as a single phase, instead decomposing 
into two coexisting phases, one Ce-rich and the other Zr-rich. 
In other words ceria-zirconia exhibits a \emph{miscibility gap}. 

However, there is uncertainty regarding the details of the phase diagram, e.g. the exact location
of the miscibility gap, and the structure of the equilibrium phases. In principle SGCMC can provide this while fully accounting for important effects such as 
cation ordering and lattice distortions \cite{Conesa2003} -- with the caveat that the interatomic potential the SGCMC is used in conjunction with is accurate. 
Here, in a first application of such methodology to metal oxide solid solutions, we use SGCMC in conjunction with the free energy method transition-matrix 
Monte Carlo (TMMC) \cite{Smith1995,Fitzgerald1999,Shen2006}, and an interatomic potential type commonly used in simulations of metal oxides 
\cite{Sayle2013,Pedone2006}, to pinpoint the miscibility gap for this system. In doing so, we add insight into the ceria-zirconia phase diagram.
%
Moreover, while bulk ceria-zirconia has received much attention, we are not aware of any theoretical studies of GB structure in this system. 
Motivated by this, we therefore also use SGCMC to calculate the equilibrium GB structures for 8 twin GBs at a wide range of temperatures and Zr concentrations in the 
Ce-rich phase. Our calculations add insight into GB structure and stability in this phase. Moreover, our calculations, and the methodology laid out
herein, lays the ground work for future miscibility gap and GB structure calculations in other metal oxide solid solutions.

\section{Results}\label{sec:results}

\subsection{Miscibility gap}
Experiment\cite{Duwez1950,Yashima1994} suggests that the miscibility gap for ceria-zirconia persists up to $\approx 2300K$. 
At 300K the Ce-rich phase has a cubic structure reminiscent of fluorite CeO$_2$, while the Zr-rich phase
takes a monoclinic structure reminiscent of ZrO$_2$. However, above $\approx$1300K the Zr-rich phase instead adopts a tetragonal fluorite-like structure.
Moreover, aside from the aforementioned cubic, tetragonal and monoclinic phases which are thought to make up the 
\emph{stable} phase diagram, various \emph{metastable} cubic/tetragonal phases have also been identified by experiment. (See \onlinecite{Devaiah2018}
for a review).
We do not explicitly consider metastable phases here; we are only concerned with the phase diagram corresponding to thermodynamic equilibrium. 
We also limit ourselves to considering the \emph{fully oxidised} ceria-zirconia system. (Further details of the phase diagram of ceria-zirconia under 
reducing conditions, which exhibits phases not found in the fully oxidised system, can be found elsewhere \cite{Devaiah2018,Conesa2003}).

Our initial aim was to calculate the miscibility gap pertaining to the cubic Ce-rich phase and the tetragonal Zr-rich phase. In other words,
our aim was to calculate the miscibility gap at high temperatures where the tetragonal, rather than the monoclinic, Zr-rich phase is stable.
Our methodology for calculating this miscibility gap is described in detail in the supplementary information. In short, it entailed using SGCMC
in conjunction with TMMC to calculate, for various temperatures, the free energy as a function of Zr concentration $c$ 
at thermodynamic conditions which correspond to the miscibility gap, $F_{\text{co}}(c)$. At such conditions, a Ce-rich and a Zr-rich phase coexist. 
In other words, two values of $c$ are equally viable in terms of thermodynamic stability. This manifests as two degenerate global minima in $F_{\text{co}}(c)$, 
one minimum at the $c$ corresponding to the Ce-rich phase, $c_1$, and the other at the $c$ corresponding to the Zr-rich phase, $c_2$. Thus from
the locations of the minima in $F_{\text{co}}(c)$, the miscibility gap concentrations $c_1$ and $c_2$ can be obtained for a given temperature.

\begin{figure}
	\includegraphics[width=\columnwidth,trim=1.5cm 1.5cm 1.5cm 2.5cm, clip]{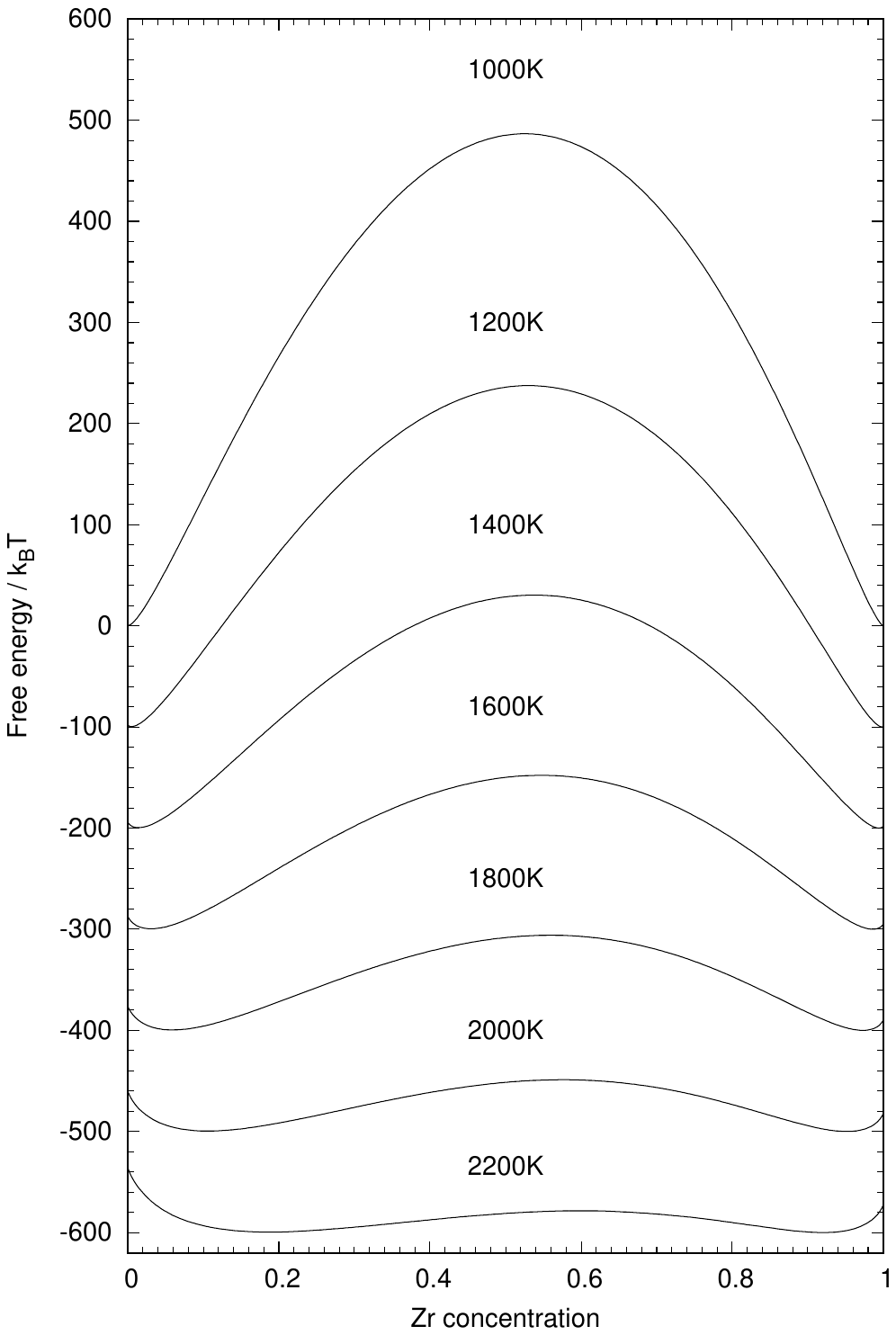}
	\caption{Free energy versus Zr concentration at the miscibility gap for bulk ceria-zirconia at selected temperatures. 
	The temperature each curve corresponds to is labelled above the curve. 
	Note that only the differences in the free energies between concentrations \emph{within} a curve are physically meaningful:
	the absolute value of the free energy is not. With this in mind, for clarity of presentation, the curves have been offset 
	from each other in gaps of -100$k_BT$, such that the global free energy minima for the 1000K, 1200K and 1400K curves are 0$k_BT$,
	-100$k_BT$ and -200$k_BT$, respectively, and similarly for the other curves}
	\label{fig:bulk_fep}
\end{figure}

Fig.~\ref{fig:bulk_fep} shows $F_{\text{co}}(c)$ obtained from our calculations for selected temperatures.
The two minima in each free energy curve correspond to the miscibility-gap Zr concentrations of the Ce-rich and
Zr-rich phases, $c_1$ and $c_2$, respectively. Note that these concentrations are very close to 0 and 1 at 1000K, and move away
from these extremes towards $c=0.5$ as the temperature is increased. This reflects the fact that the solubility of Zr in the Ce-rich phase increases 
with temperature, and similarly for Ce in the Zr-rich phase.
Note also that the size of the free energy barrier at $c\approx 0.5$ decreases as temperature is increased. 
This is expected; the free energy barrier becomes smaller as the temperature is increased, in principle vanishing at the critical (eutectoid) temperature.

\begin{figure}
	\includegraphics[width=\columnwidth,trim=1.5cm 1.5cm 1.5cm 1.5cm, clip]{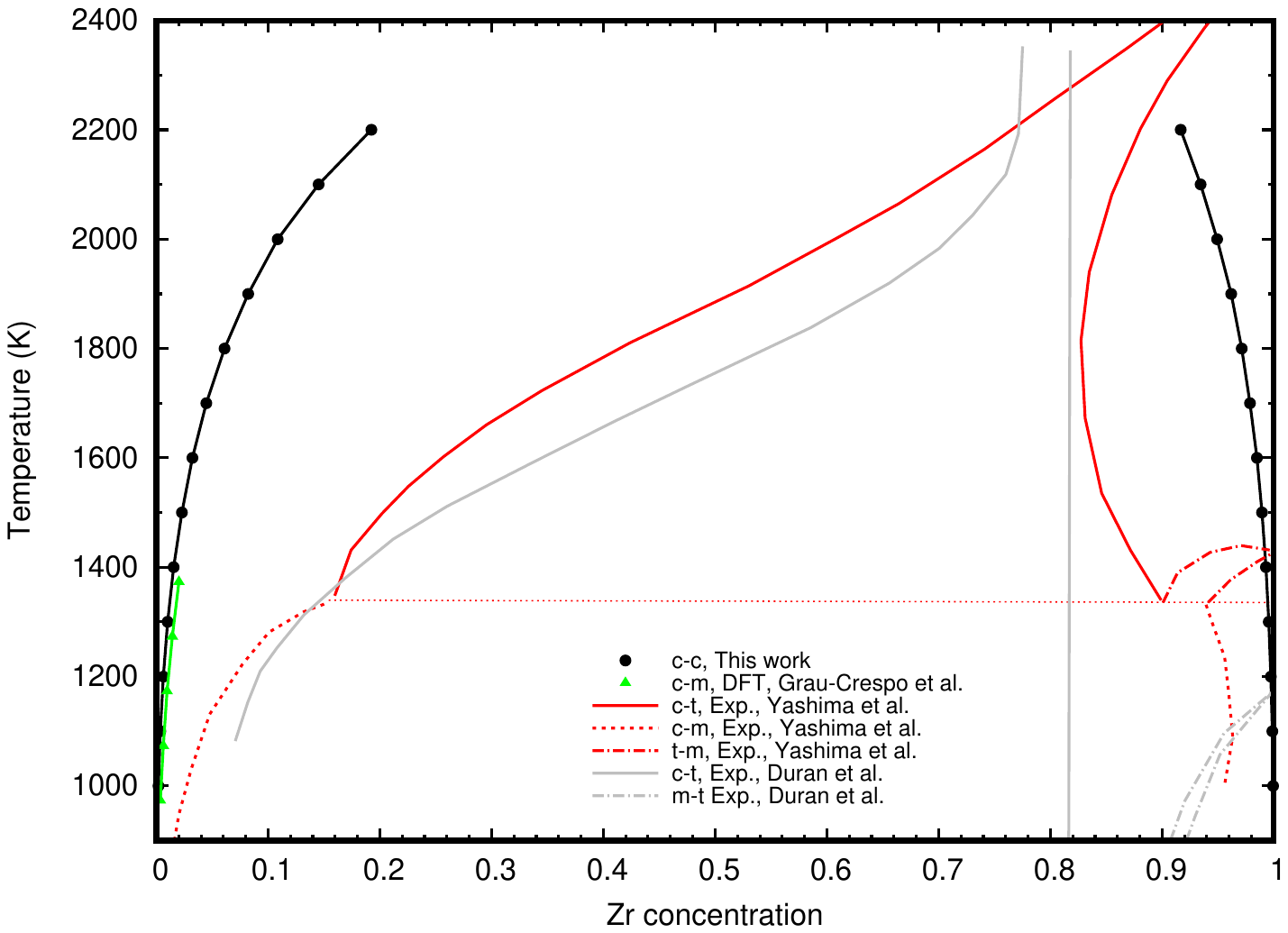}
	\caption{Phase boundaries for ceria-zirconia. Here `c', `t' and `m' denote the cubic, tetragonal and monoclinic phases, respectively.
             Moreover, `c-t' denotes a phase boundary pertaining to the cubic Ce-rich phase and the tetragonal Zr-rich phase. Note that for
             each phase boundary there are two curves; the miscibility gap is the region enclosed by the curves.
             Data are shown from this work, the DFT study by Grau-Crespo et al. \cite{Grau-Crespo2011}, and the experimental studies by
             Duran et al. \cite{Duran1990} and Yashima et al. \cite{Yashima1994,Zhang2006}.}
	\label{fig:bulk_phase_diagram}
\end{figure}

The miscibility-gap compositions $c_1$ and $c_2$ obtained from the $F_{\text{co}}(c)$ for all temperatures we considered are
plotted in Fig.~\ref{fig:bulk_phase_diagram}. 
From Fig.~\ref{fig:bulk_phase_diagram} it can be seen that the miscibility gap obtained from our calculations is not symmetrical about $c=0.5$, 
rather it is skewed towards $c=1$. This reflects the fact that Zr is more soluble in the Ce-rich phase than Ce is in the Zr-rich phase, something
which can be explained by the fact that Zr cations are smaller than Ce cations.

\subsubsection{Comparison with other studies}
While in reality, as mentioned above, the Zr-rich phase takes a tetragonal crystal structure, this is not borne out in our calculations. Rather,
our calculations yielded a \emph{cubic} structure for the Zr-rich phase: we did not observe any significant deviations from a cubic 
structure in the Zr-rich phase in our simulations. Thus Fig.~\ref{fig:bulk_fep} and the data pertaining to this work in
Fig.~\ref{fig:bulk_phase_diagram} describe a miscibility gap pertaining to a cubic Ce-rich and a \emph{cubic} Zr-rich phase. This is due to the 
shortcomings of our choice of interatomic potential. It is known that interatomic potentials of the type that we used, namely pair potentials, have limitations
in ceria-zirconia \cite{Conesa2003}. In particular, they cannot capture the more covalent nature of the Zr--O bonds. 
In the supplementary material we show that the model gives reasonable agreement with experiment for various 
properties of cubic ceria-zirconia at 300K. Hence it may be the case that this accuracy of the model breaks down as temperature is increased or at
very high Zr concentrations. 
We discuss this issue further in Section~\ref{sec:conclusions}.

Also shown in Fig.~\ref{fig:bulk_phase_diagram} are miscibility gaps obtained from DFT \cite{Grau-Crespo2011} and
experiment \cite{Duran1990,Yashima1994}. 
Note that the DFT miscibility gap pertains to thermodynamic equilibrium between the cubic Ce-rich 
phase and the \emph{monoclinic} Zr-rich phase rather than the tetragonal phase, which may explain any discrepancies between the DFT data and the results 
of this work. Further possible reasons for discrepancies are provided in Ref.~\onlinecite{Grau-Crespo2011}. 
Note that only the component of the miscibility gap for the Ce-rich phase was provided in Ref.~\onlinecite{Grau-Crespo2011}, which is why the DFT data
in the figure has only one curve rather than two.
Note also that the DFT calculations employed various approximations, in particular a very small system size, an on-lattice structural model, and an 
ideal-solution approximation for the entropy of mixing. 
By contrast, while our calculations employ an interatomic potential which is not as accurate a description of the interatomic interactions as DFT, 
our calculations are more accurate in all other regards: they employ a much larger system size (1500 atoms); capture lattice distortions (including anharmonic
vibrations) and variations in volume with concentration; and in our calculations of free energies the entropy is in effect treated exactly.

It can be seen from Fig.~\ref{fig:bulk_phase_diagram} that our miscibility gap is in good agreement with the the DFT study. 
However, comparing our results to those of experiment, it can be seen that our model significantly underestimates the solubilities of Zr and Ce in, 
respectively, 
the Ce-rich and Zr-rich phases. In other words our miscibility gap is much larger than the experimental miscibility gap.
This could be attributed to limitations of the interatomic potential discussed above which leads to the Zr-rich phase exhibited by the
model being cubic instead of tetragonal. 
For instance, one particular limitation of the model is that the cation charges are fixed; it is known that, the cation
charges, in particular those of Ce, are environment-dependent.\cite{Conesa2003}
However, the DFT results also massively underestimate the solubility of Zr in Ce.
With this in mind, as as been suggested previously\cite{Grau-Crespo2011}, another possibility is that high cation diffusion barriers in experiment prevent 
the solid solutions from reaching thermodynamic equilibrium in experimental timescales. This would result in experiment underestimating the size of the 
miscibility gap, or, equivalently, overestimating in the solubilities of Zr in the Ce-rich phase and Ce in the Zr-rich phase.
Note also that the DFT and our results pertain to ceria-zirconia in its \emph{fully-oxidised} state. However, this state
might not be precisely realised in experiment; the phase diagram of reduced ceria-zirconia is known to differ 
from that of fully-oxidised ceria-zirconia \cite{Devaiah2018,Conesa2003}.

Nevertheless, our calculations reproduce key qualitative features of the ceria-zirconia phase diagram. Firstly, we predict the existence of a miscibility
gap in a comparable temperature range to that predicted by experiment; it is noteworthy this is not guaranteed for a given choice of interatomic potential 
for this system. Secondly, we recover the correct crystal structure, namely the cubic fluorite phase, for the Ce-rich phase. Given that the focus of our 
GB simulations is the Ce-rich phase, we hence believe that our calculations will give at least qualitatively accurate predictions.

\subsection{Grain boundary structure}
We now turn to GBs. We consider 8 twin GBs, namely $\Sigma 3(111)$, $\Sigma 3(211)$, $\Sigma 5(210)$, $\Sigma 5(310)$, $\Sigma 9(221)$, $\Sigma 13(320)$,
$\Sigma 13(510)$ and $\Sigma 17(410)$. 
We limit ourselves to the Ce-rich phase only, and, again, consider temperatures ranging from 1000K to 2200K. Moreover we consider bulk Zr compositions,
$c$, ranging from the impurity limit up to $c_1$, which recall is the Zr composition for the Ce-rich phase at the miscibility
gap or, in other words, the limit of solubility for Zr in the Ce-rich phase for our choice of model.

\subsubsection{General behaviour: Zr segregates to grain boundaries}
\begin{figure*}
  \subfloat[]{
    \includegraphics[width=0.95\columnwidth,trim = 1.5cm 2.5cm 1.5cm 2.5cm, clip]{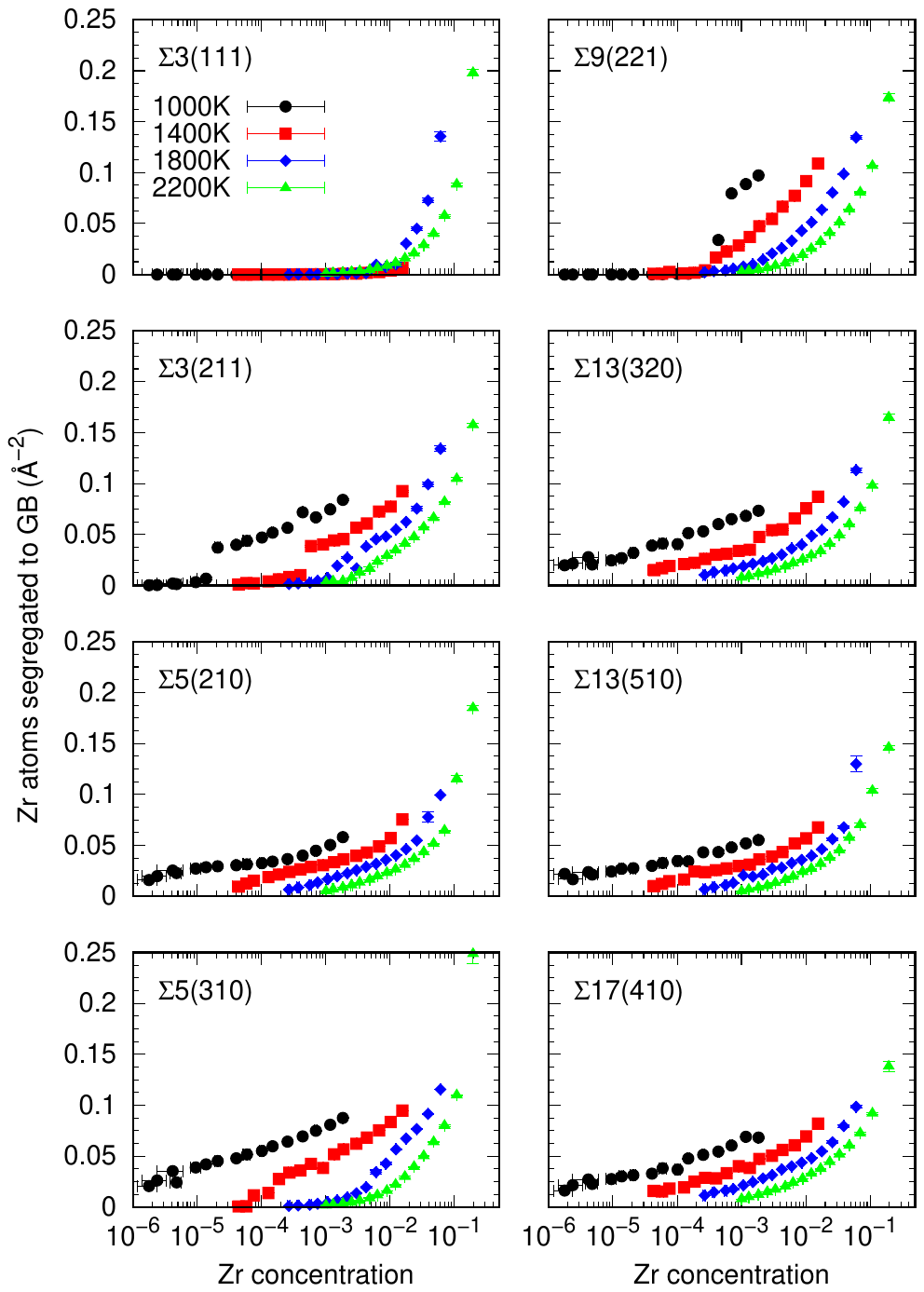}
	\label{fig:segregation_metric_Ce_rich}
  }
  \qquad
  \subfloat[]{
    \includegraphics[width=0.95\columnwidth,trim = 1.5cm 2.5cm 1.5cm 2.5cm, clip]{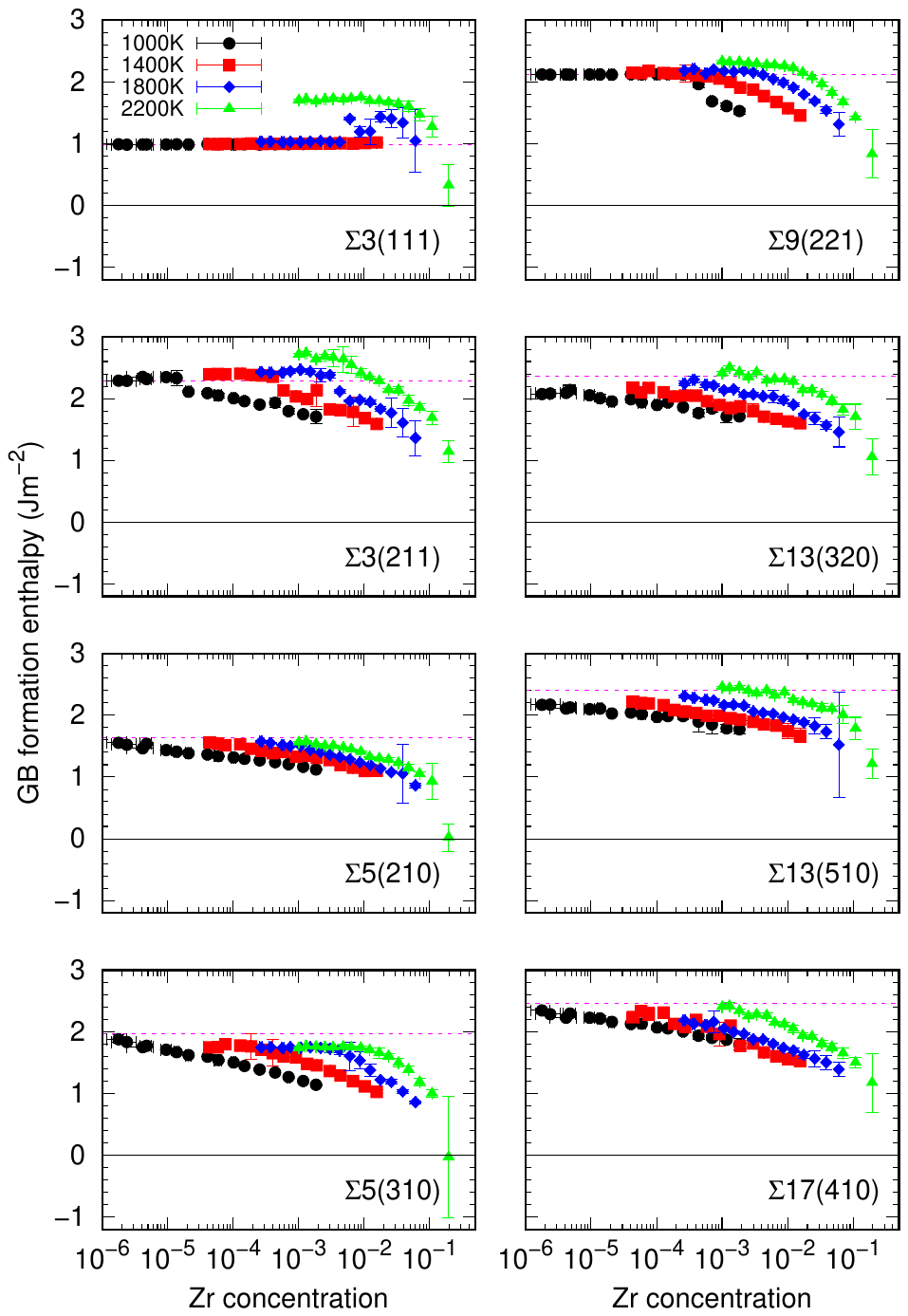}
	\label{fig:formation_enthalpies}
  }
 \caption{(a) Amount of Zr segregation (i.e. the \emph{excess} number of Zr per unit area of GB, $\xi$, as defined in Eqn.~\ref{eqn:seg_metric}), and
    (b) GB formation enthalpies (see Eqn.~\ref{eqn:GB_formation_enthalpy}),
    for various GBs in the Ce-rich phase of fluorite ceria-zirconia, at various temperatures and bulk Zr concentrations. 
    Note that for each temperature the highest Zr concentration corresponds to the miscibility-gap concentration for that temperature, $c_1$. 
    Also shown in (b) for comparison for each GB is the formation enthalpy for \emph{pure} ceria at 1000K, which is given by a dashed magenta line.} 
\end{figure*}

We begin by considering the extent of segregation of Zr to the GBs. To characterise the segregation we use a quantity $\xi$ defined in
the supplementary information (Eqn.~\ref{eqn:seg_metric}) which, loosely speaking, is the number of Zr atoms per unit area of GB which is in \emph{excess} of 
the amount expected given a random distribution of Zr and the $c$ under consideration.
Fig.~\ref{fig:segregation_metric_Ce_rich} shows the amount of Zr segregation, i.e. $\xi$, versus $c$ for the GBs we considered at the temperatures we considered.
It can be seen from the figure that, at a fixed temperature, the amount of Zr segregation to the GBs
\emph{in general} increases with $c$ -- with the maximal segregation for a given temperature occurring at
the miscibility gap concentration $c_1$. (The special case of the $\Sigma 3(111)$ GB is discussed in a moment).
Moreover, in general, at a given $c$, the segregation is more pronounced at \emph{lower} temperatures. 
This is clear in the data for the $\Sigma 5(210)$, $\Sigma 5(310)$, $\Sigma 13(320)$, $\Sigma 13(510)$ and $\Sigma 17(410)$ 
GBs. 

\subsubsection{Grain-boundary--specific segregation behaviour}
Fig.~\ref{fig:segregation_metric_Ce_rich} also reveals that the amount of Zr segregation to a GB is 
sensitive to the GB structure. Moreover, there are \emph{qualitative} differences in the segregation behaviour
between GBs.
To elaborate, while the $\Sigma 5(210)$, $\Sigma 5(310)$, $\Sigma 13(320)$, $\Sigma 13(510)$ and 
$\Sigma 17(410)$ GBs exhibit qualitatively similar behaviour, 
with $\xi$ varying continuously with $c$ for all concentrations we probed, the 
$\Sigma 3(111)$, $\Sigma 3(211)$ and $\Sigma 9(221)$ GBs exhibit different behaviour:
\begin{itemize}
\item $\Sigma 3(111)$ exhibits essentially no segregation at 1000K and 1400K, i.e. the GB is \emph{clean}.
However, at 1800K segregation analogous to what is observed in the other GBs occurs as $c$ approaches the
the miscibility-gap composition $c_1$. At 2200K the behaviour is similar to the other GBs.
\item $\Sigma 3(211)$ is clean for $c\lessapprox 10^{-5}$ for all temperatures.
However, at 1000K and $c\approx 10^{-5}$ there is a discontinuous 
jump to $\xi\approx 0.04$. Similar is observed at 1400K and $c\approx 10^{-3}$, though the
jump in $\xi$ is smaller. 
\item $\Sigma 9(221)$ similarly is clean for $c\lessapprox 10^{-4}$. However, there
is a discontinuous increase in $\xi$ at 1000K and $c\approx 10^{-3}$ which is much more 
pronounced than that in $\Sigma 3(211)$. A smaller discontinuity occurs at a slightly higher
$c$ at 1400K.
\end{itemize}

\subsubsection{First-order complexion transitions}
While physical properties of a GB (e.g. the amount of segregated dopant atoms at the GB, or the GB formation energy) 
typically change continuously with thermodynamic parameters such as temperature and dopant concentration, it is also possible for 
discontinuous changes to occur at certain special thermodynamic parameters \cite{Cantwell2020}. Such discontinuities are akin to 
first-order phase transitions commonly observed in bulk systems; they are a result of first-order transitions between GB 
\emph{complexions} -- where the term `complexion' has become standard to describe an interfacial analogue of a bulk phase whose 
structure is distinct from any phases found in the bulk \cite{Dillon2007,Cantwell2014,Harmer2016,Kaplan2013,Cantwell2020}.
With this in mind, the discontinuities mentioned above in the amount of Zr segregation versus $c$ in the $\Sigma 3(211)$ 
and $\Sigma 9(221)$ suggest the existence of first-order complexion transitions in these GBs.
Interestingly, however, in these GBs there are no discontinuities at 1800K and 2200K.
This, in combination with the fact that the jumps in the Zr segregation at the discontinuities are smaller at 1400K
than 1000K, suggest that the first-order complexion transitions terminate at critical points located
between 1400K and 1800K. From examining the $c$ at which the discontinuities in the Zr segregation occur,
we speculate that the critical concentration is at $c\approx 10^{-3}$ for the 
$\Sigma 3(211)$ GB and $c\approx 5\times 10^{-4}$ for the $\Sigma 9(221)$ GB.

\begin{figure}
	\includegraphics[width=\columnwidth]{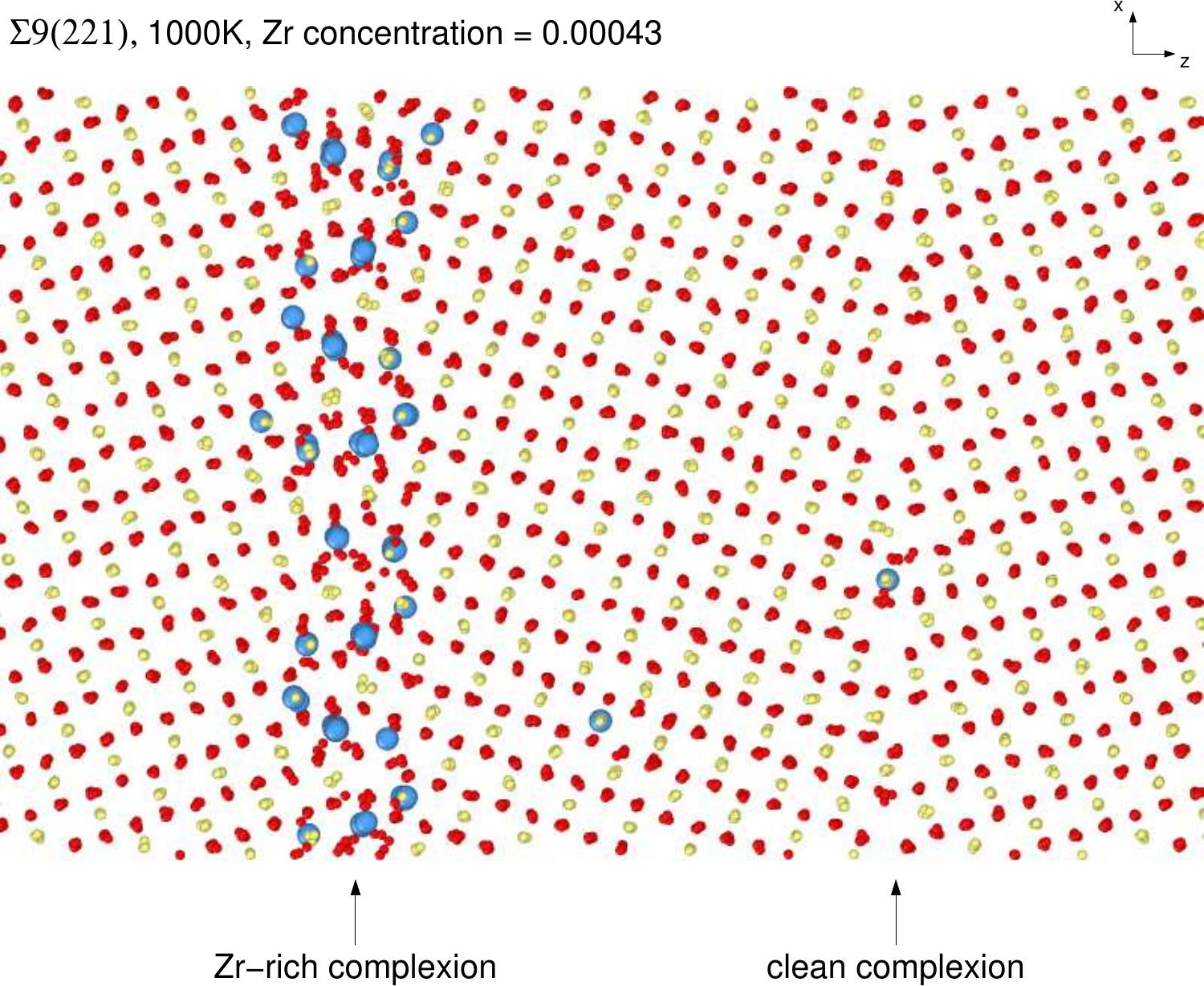}
	\caption{A configuration obtained from our simulation of the $\Sigma 9(221)$ GB at 1000K and $c=0.00043$, exhibiting coexistence of a
	clean complexion and a Zr-rich complexion. In this figure Ce atoms are coloured 
	light yellow, O atoms are coloured red, and Zr atoms are coloured blue. Moreover
	Zr atoms have been given a larger radius than O and Ce atoms for emphasis. The
	plane of the figure is the $x/z$ plane, where the $z$ direction is chosen to be perpendicular to the GB.}
	\label{fig:snapshot_S9_221_1000K_coexistence}
\end{figure}

Further evidence for the first-order nature of these transitions can be found in our observation of two complexions, corresponding
to the complexions either side of the transition, \emph{coexisting} in a single simulation. To elaborate, our simulation
cells each contained two GBs. We observed in our simulation for the $\Sigma 9(221)$ GB
at 1000K and $c=0.00043$ that the two GBs in the simulation cell had
different structures.  A configuration from that simulation is shown in
Fig.~\ref{fig:snapshot_S9_221_1000K_coexistence}, where it can be seen that one of the GBs
takes the `clean' complexion corresponding to no Zr segregation, which is stable for 
$c\lessapprox 0.00043$, while the other GB takes the `Zr-rich' complexion
which is stable for $c\gtrapprox 0.00043$. 
This suggests that there is a free energy barrier which prevents a clean 
complexion from spontaneously transforming into a Zr-rich complexion (or vice versa), even if such a change were to minimise 
the free energy of the system as a whole. Such behaviour is a signature of a first-order transition.

In principle, free energy methods could be used to \emph{prove} that the transition is first-order.  Plots of free energy versus Zr 
concentration in Fig.~\ref{fig:bulk_fep} show the existence of a free energy barrier separating the Ce-rich and Zr-rich phases in 
bulk ceria-zirconia. Similarly, at thermodynamic conditions close to a complexion transition, by calculating the free energy versus 
an appropriate order parameter for the transition, e.g. the quantity of Zr segregated to the GB, the existence of a free energy barrier could be
demonstrated.

\subsubsection{Multisite segregation mechanism}
We now consider the positional distribution of segregated Zr atoms at the GBs.
For the sake of brevity here we do not discuss all GBs and thermodynamic conditions in depth, instead focusing on selected GBs.

In accordance with continuous variation of the Zr segregation with $c$ as discussed above, in most of the GBs
and thermodynamic conditions we considered the Zr positional distribution at a GB varies continuously as 
$c$ is increased.
Moreover, analysis of configurations obtained from our simulations reveals that the mechanism of Zr segregation as $c$ is increased
in such GBs is akin to multisite adsorption. To elaborate, as $c$ is increased, the probabilities of cation sites at the GB 
being occupied by a Zr atom also increases, but with rates for different sets of symmetrically equivalent cation sites.
This behaviour is illustrated in Figs.~\ref{fig:snapshots_S5_210_1000K} and \ref{fig:snapshots_S5_310_1000K}, which show the Zr occupancy 
of cation sites in the $\Sigma$5(210) and $\Sigma$5(310) GBs, respectively, as $c$ is varied at 1000K. In both
figures the highest $c$ considered corresponds to the miscibility-gap composition for 1000K. 
The multisite nature of the segregation as Zr concentration is increased at 1000K in these GBs is clear from these figures.

\subsubsection{Symmetry breaking}
\begin{figure*}
  \subfloat[]{
	\includegraphics[width=0.9\columnwidth]{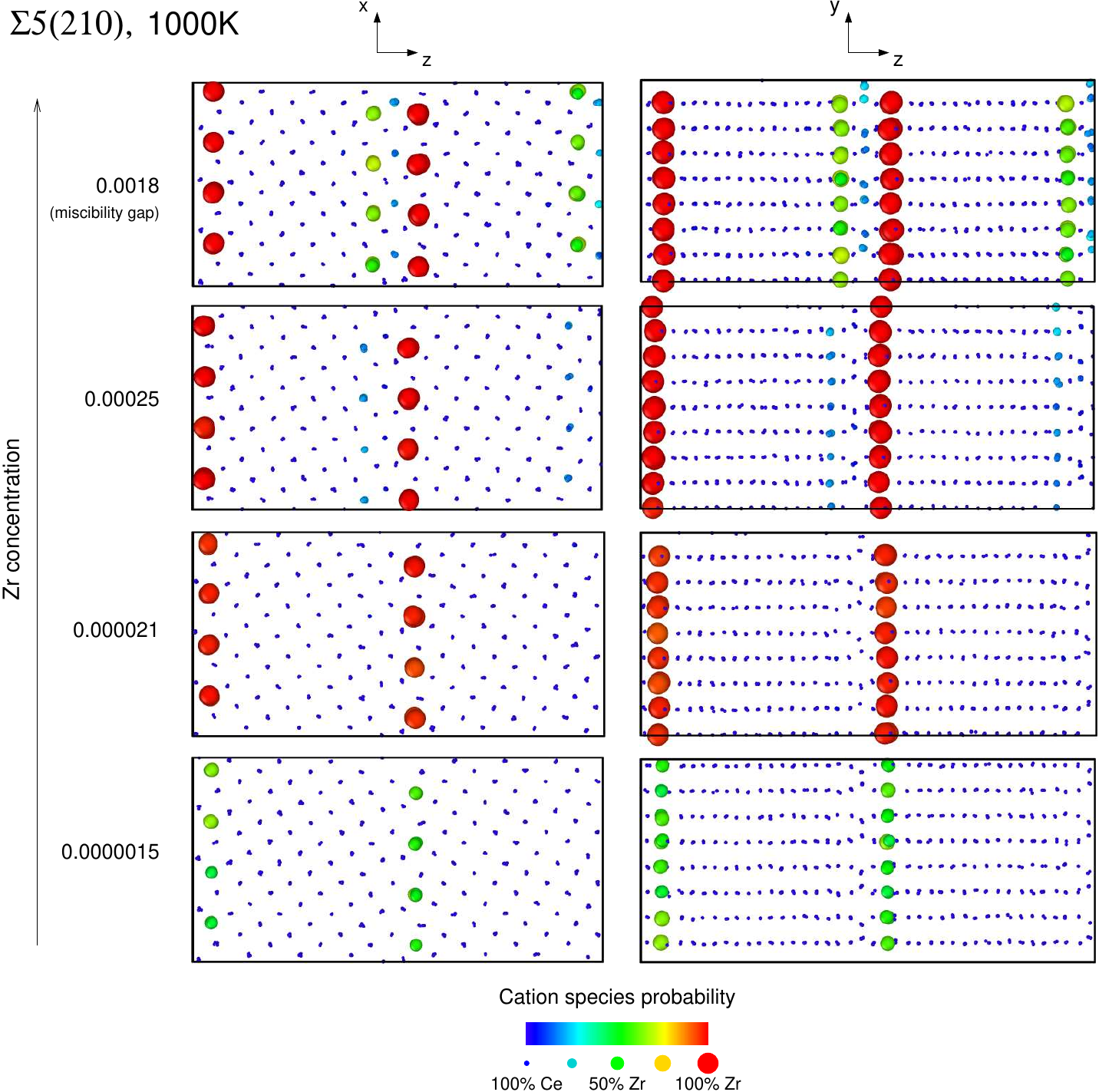}
	\label{fig:snapshots_S5_210_1000K}
  }
  \qquad
  \subfloat[]{
    \includegraphics[width=0.9\columnwidth]{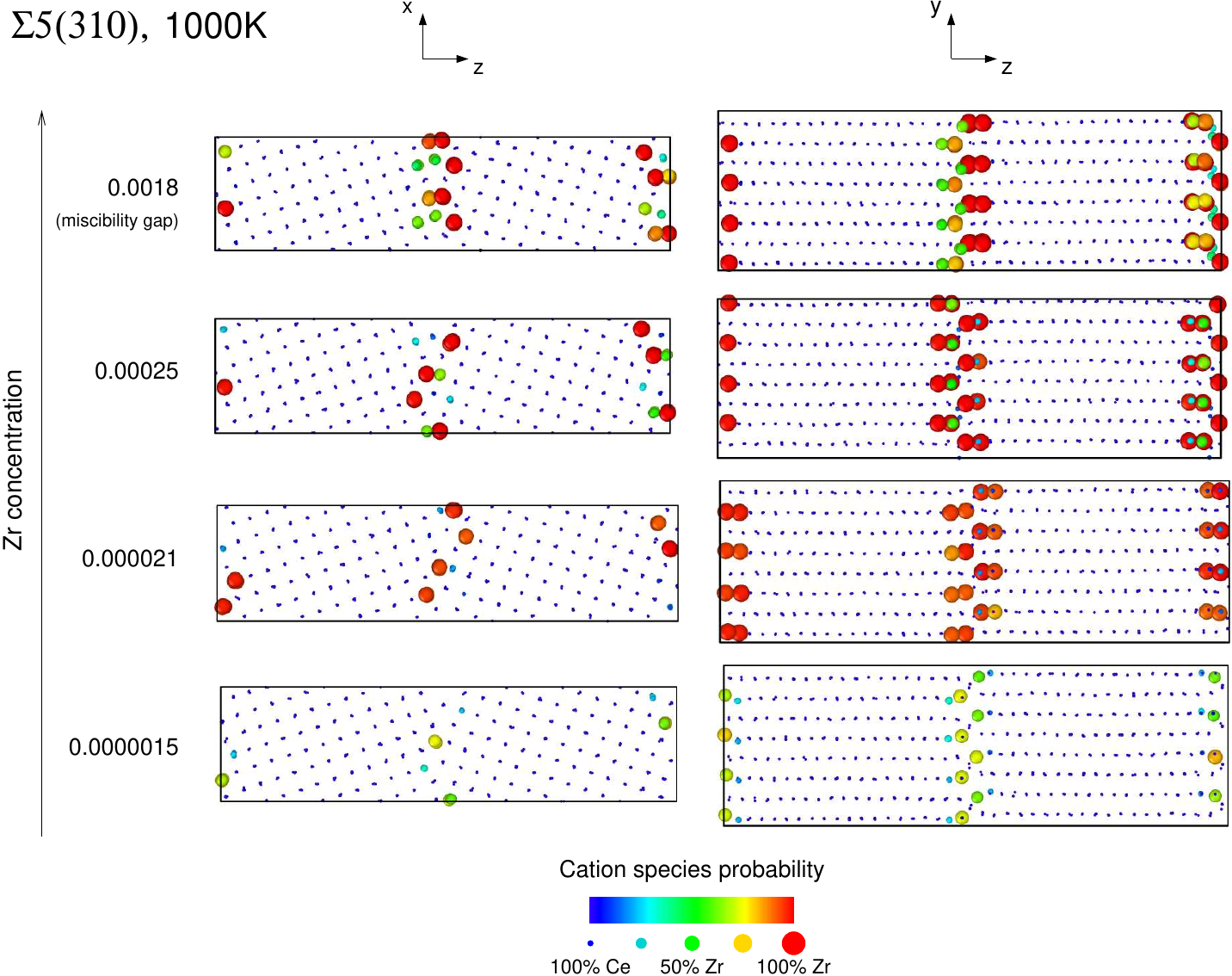}
	\label{fig:snapshots_S5_310_1000K}
  }
  \caption{Cation positions and occupancy (Ce versus Zr) for the $\Sigma$5(210) and $\Sigma$5(310) GBs obtained from simulations at 1000K and various 
    bulk Zr concentrations. (a) corresponds to the $\Sigma$5(210) GB. Here, each row corresponds to a certain Zr concentration, as labeled. Moreover, for each 
    concentration the left image shows the GB structure in the $x/z$ plane while the right image shows the same structure in the $y/z$ plane, where the $z$ 
    direction is chosen to be perpendicular to the GB. In each image the cations are represented by circles, whose positions are taken from a 
    representative configuration 
    from a simulation trajectory. The size and colour of the circles represents the probability of the site being Ce or Zr, based on the amount of time the site spent as, 
    e.g. Ce, over a long simulation. Smaller circles with colder colours correspond to a higher probability of the site being Ce, while larger circles with warmer colours 
    correspond to a higher probability of the site being Zr. (b) is similar but for the$\Sigma$5(310) GB.}
\end{figure*}

These figures also illustrate another interesting feature of some of the GB structures we observed, namely \emph{symmetry breaking}.
To elaborate, by definition twin GBs, for pure structures, have reflection symmetry with respect to the plane which constitutes the
GB. By contrast, we observed GB structures in ceria-zirconia where this is not the case. The $\Sigma 5(210)$ GB 
is an especially striking example of this, as can be seen in Fig.~\ref{fig:snapshots_S5_210_1000K}. Note that the Zr atoms preferentially segregate 
to one side of the GB. For instance, at $c=0.0000015$ the segregation sites correspond to a monolayer which 
is off-centre with regards to the location of the GB, while the symmetrically equivalent monolayer on the other side of the 
GB does not exhibit any Zr segregation. 

In the supplementary information we provide the density of Zr versus distance perpendicular to the GB for all the GBs and conditions
we considered (Fig.~\ref{fig:zdensity}). As well as providing a more quantitative measure of GB structure than figures such as Figs.~\ref{fig:snapshots_S5_210_1000K} and 
\ref{fig:snapshots_S5_310_1000K} can provide, Fig.~\ref{fig:zdensity} also shows the symmetry breaking in the $\Sigma 5(210)$ GB from a different
perspective. Moreover, this figure reveals that symmetry breaking is present in other GBs, e.g. the $\Sigma 13(320)$ GB.

Interestingly, the $\Sigma 5(210)$ GB, while exhibiting symmetry breaking up to 1800K, does not appear to do so at 2200K. This can be seen in
Fig.~\ref{fig:snapshots_MG_2200K}, which shows the cation occupancy of selected GBs at the miscibility-gap composition at 2200K. From this figure
it appears that the $\Sigma 5(210)$ GB has a symmetrical GB structure at these conditions.

\subsubsection{Sites exhibiting Ce segregation}
\begin{figure*}
	\includegraphics[width=1.8\columnwidth]{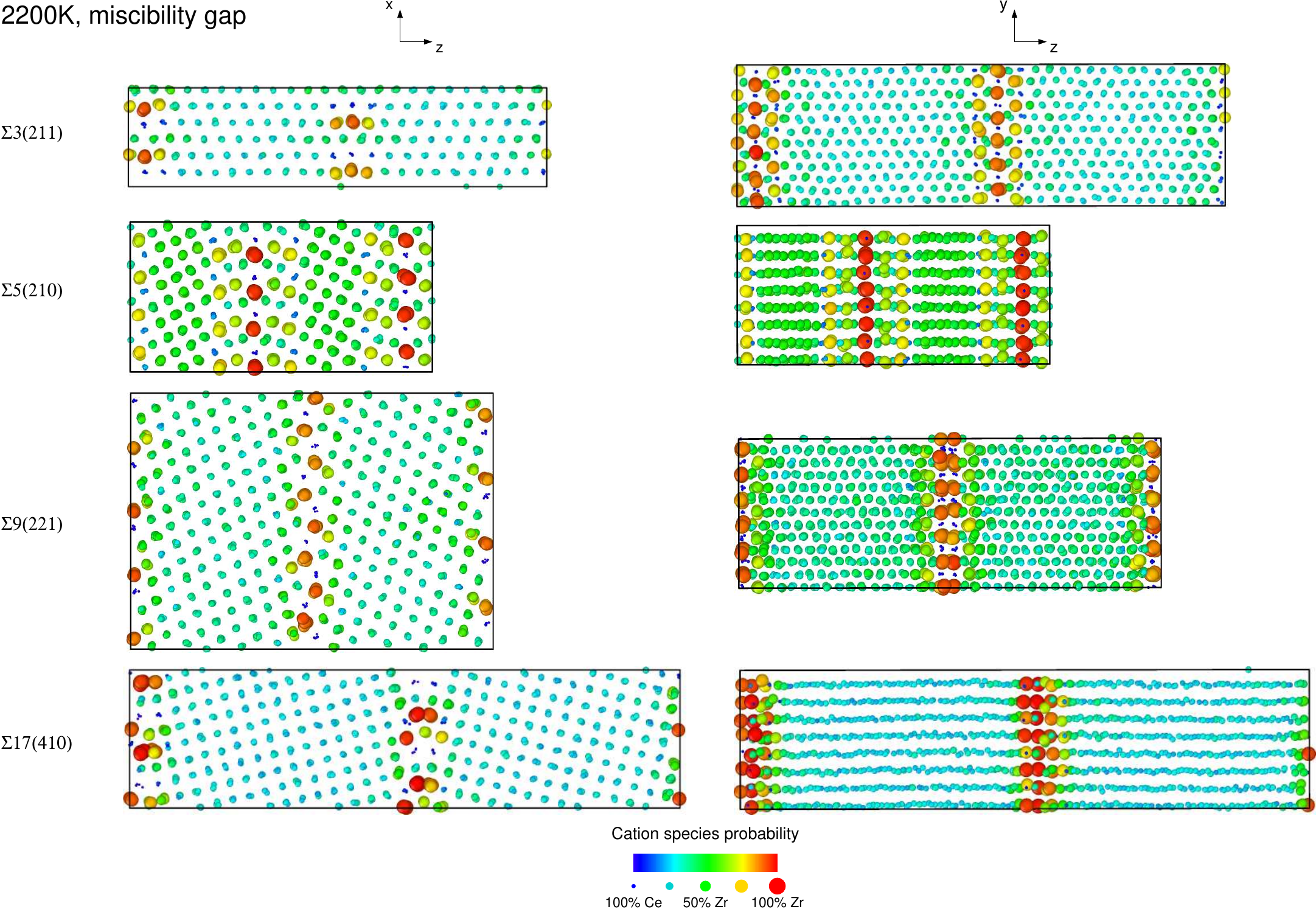}
	\caption{Cation positions and occupancy (Ce versus Zr) for selected GBs obtained from simulations at 2200K and the miscibility-gap Zr concentration. The 
	details of this figure are similar to that of Fig.~\ref{fig:snapshots_S5_210_1000K}. }
	\label{fig:snapshots_MG_2200K}
\end{figure*}

Fig.~\ref{fig:snapshots_MG_2200K} also highlights another interesting structural feature present in some GBs, namely, that
some cation sites at the GB are occupied by Zr with a \emph{lower} probability than in the bulk. In other words, such sites exhibit Ce segregation,
as opposed to Zr segregation. This can be seen clearly in the figure: for all the GBs shown in the figure, moving along the GB 
in the $x$ direction, one encounters cation sites which alternate in whether they exhibit Zr or Ce segregation.

\subsection{Grain boundary stability}
Having discussed features of their equilibrium structures, we now turn to quantifying the \emph{stability} of the GBs.
To characterise the stability of a GB we calculated the enthalpy of formation per unit area of the GB at the $c$ and temperature
under consideration. Details of how we calculated this quantity are provided in the supplementary information (Section~\ref{sec:method}, and the relevant equation is
Eqn.~\ref{eqn:GB_formation_enthalpy}). Note that the enthalpy of formation we consider includes the enthalpy change due to segregation:
it is the enthalpy associated with creating an \emph{equilibrium} GB structure including any segregation within the solid solution, 
starting from the bulk solid solution.

Fig.~\ref{fig:formation_enthalpies} shows the formation enthalpies for the GBs, $c$ and temperatures we considered. 
For comparison, the 1000K formation enthalpies of the \emph{pure} ceria GBs, 
i.e. $c=0$, are also shown in this figure. (Moreover Fig.~\ref{fig:pure_GB_formation_enthalpies} in the supplementary information
provides the formation enthalpies for pure ceria GBs at the other temperatures we considered).
From Fig.~\ref{fig:formation_enthalpies} it can be seen that as $c$ is increased from 0, for almost all of the GBs the formation enthalpy for the GB decreases
at a fixed temperature. Comparing this figure to Fig.~\ref{fig:segregation_metric_Ce_rich}, it can be seen that such decreases in the formation enthalpy 
track increases in the Zr segregation $\xi$ to the GB with $c$, e.g. exhibiting discontinuous jumps at certain $c$ where there are such jumps in $\xi$ . 
This suggests that \emph{Zr segregation stabilises the GBs}. Moreover, \emph{doping ceria with Zr stabilises the GBs}. The exception is the $\Sigma 3(111)$ GB 
at 1000K--1800K, which, as discussed earlier, exhibits no Zr segregation and hence does not become more stable as a result of Zr doping.

\begin{figure}
	\includegraphics[width=\columnwidth,trim=1.5cm 1.5cm 1.5cm 1.5cm, clip]{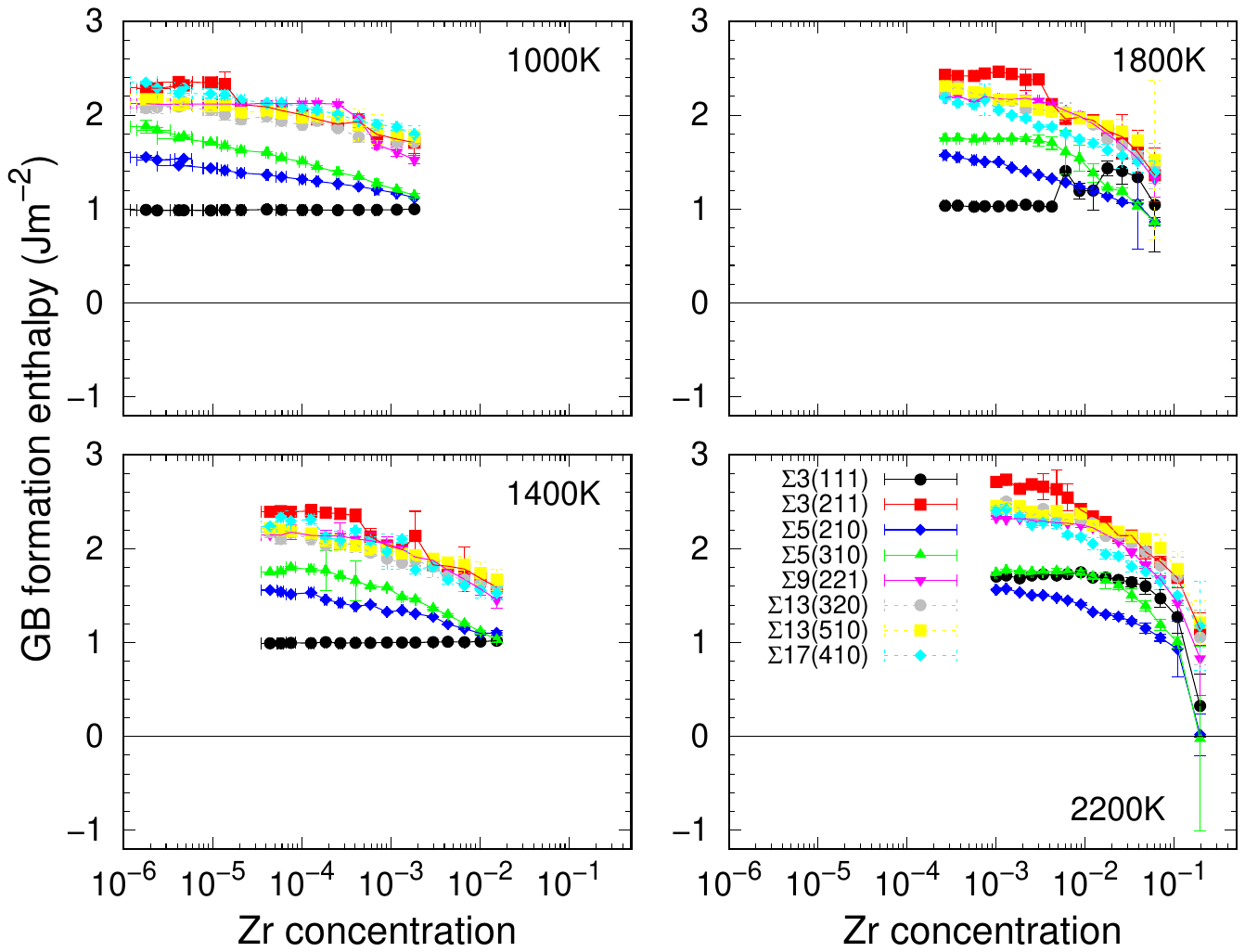}
	\caption{GB formation enthalpies  (see Eqn.~\ref{eqn:GB_formation_enthalpy}) for various GBs in the Ce-rich phase of fluorite ceria-zirconia, 
    at various temperatures and bulk Zr concentrations -- comparing the relative stabilities of the GBs	at each Zr concentration and temperature.
	For each temperature the highest Zr concentration corresponds to the miscibility-gap concentration for that temperature.} 
	\label{fig:formation_enthalpies_2}
\end{figure}

As well as stabilising the GBs, varying $c$ also affects the stability of the GBs \emph{relative} to each other. 
This can be seen in Fig.~\ref{fig:formation_enthalpies_2}, which presents the same data as shown in Fig.~\ref{fig:formation_enthalpies}, but enables 
the stabilities of the GBs relative to each other to be discerned more easily. 
From this figure it can be seen that for 1000K--1400K the effect of increasing $c$ is to bring the formation enthalpies
of all GBs closer to that of the $\Sigma 3(111)$, which is the most stable at $c=0$. In fact, at the highest value of $c$ considered, 
which recall is the miscibility-gap composition, the $\Sigma 5(210)$ and 
$\Sigma 5(310)$ have comparable energies to the $\Sigma 3(111)$ GB at 1000K--1400K. At 2200K these three
GBs have comparable energies (see Fig.~\ref{fig:pure_GB_formation_enthalpies} in the supplementary information) in pure ceria, 
but the effect of doping with Zr is to make the $\Sigma 5(210)$ significantly more stable than the other two -- and hence the most stable GB -- 
for most of the $c$ we considered. Doping of ceria with Zr has therefore affected the hierarchy of stability of different GB types.

\subsubsection{Implications for microstructure}
GB formation enthalpies play an important role in the population and growth of different GB types in a material; GBs with lower formation enthalpies are 
expected to be more prevalent in the material. With this in mind, our data suggest that doping with Zr would increase the diversity of GBs in ceria-zirconia 
at 1000K--1800K. To elaborate, at 1000K--1800K in ceria the $\Sigma 3(111)$ GB is expected to be the most common on account of the fact it has a significantly lower 
formation enthalpy than other GB types. However, as mentioned above, at these temperatures increasing the Zr concentration lowers the formation enthalpies of 
other GB types such that they become comparable to that of the $\Sigma 3(111)$ GB, the result being that they should be more prevalent in the system.

At 2200K the situation is different. Recall that for 1000K--1800K the formation enthalpy of the $\Sigma 3(111)$ GB is essentially unaffected by Zr concentration.
Moreover, this GB always has the lowest formation enthalpy at these temperatures (see Fig.~\ref{fig:formation_enthalpies_2}). 
Hence the \emph{lowest} enthalpy required to create a certain area of \emph{any} GB is not affected by Zr concentration for 1000K-1800K. 
By contrast, as can be seen from Fig.~\ref{fig:formation_enthalpies_2}, at 2200K it is the $\Sigma 5(210)$ GB which has the lowest formation enthalpy of all the
GB types we considered, and increasing the Zr concentration \emph{does} affect this enthalpy, lowering it. Hence at 2200K
the enthalpy required to create \emph{any} GB decreases with increasing Zr concentration. For this reason we might expect that doping with
Zr increases the propensity of the system to be polycrystalline, i.e. to have smaller grain sizes, at this temperature.

\section{Discussion}\label{sec:conclusions}
\subsection{Phase separation}
We have applied the semi-grand canonical Monte Carlo (SGCMC) method to the metal oxide solid solution ceria-zirconia. Firstly, we demonstrated that,
in conjunction with free energy methods, it is possible to use SGCMC to calculate precise miscibility gaps for interatomic potentials of the type
routinely used in atomistic simulations of metal oxides. We demonstrated this by calculating the miscibility gap for ceria-zirconia. Moreover,
we compared our predicted miscibility gap to experiment and DFT--based results. 
Agreement between our calculations and DFT was good. However, agreement with experiment was merely
qualitative. (Note, however, that the DFT--based theoretical results were in comparably poor agreement with experiment).

We discussed possible reasons for this, one of which is inadequacies in the interatomic potential model.  Such models are typically fitted to low 
temperature bulk properties, and, given the well-known issue with \emph{transferability} of interatomic potentials to thermodynamic conditions far
from where they were optimised, it is optimistic to expect that our choice of potential could yield an accurate high-temperature phase diagram. 
Noting that our methodology is in principle exact for a given choice of model for interatomic interactions in the sense that it samples, e.g. lattice 
distortions and 
different cation arrangements which manifest at high temperatures, an obvious solution to this problem is to use a more accurate interatomic potential. 
In principle our methodology could be used with a DFT-based model for the interatomic interactions. However, in practice this
is intractable on account of the considerable computational cost involved. Alternatively, our methodology could be employed with emerging classes of 
interatomic potential \cite{Behler2016} which approach DFT-level accuracy with considerably less computational cost than direct DFT.
Another intriguing future application of our methodology is to aid the development of new interatomic potentials which give
closer matches to experimental miscibility gaps. There is precedent for this; accurate methods for pinpointing fluid \cite{Martin1998} and solid-solid 
\cite{Mendelev2016} 
phase transitions have been used in other contexts to develop well-used interatomic potentials which yield realistic locations of phase transitions.

\subsection{Grain boundary structure and stability}
We also used SGCMC to determine equilibrium structures of various twin grain boundaries (GBs) in the Ce-rich phase of ceria-zirconia for a range of temperatures 
and Zr concentrations. Our calculations predict an interesting richness in the equilibrium GB structures in this system. Our key predictions are
summarised below:
\begin{enumerate}
    \item In general, Zr segregates to GBs in the Ce-rich phase, with the Zr segregation proceeding continuously as the Zr concentration is increased via a mechanism 
    akin to multisite adsorption.
    \item At certain thermodynamic conditions and certain GBs there is \emph{no} discernible segregation. The $\Sigma 3(111)$ GB, notable because it
    is the most stable twin GB in pure ceria up to $\approx$1800K, is one such GB, only exhibiting appreciable segregation at
    compositions close to the miscibility gap for temperatures above 1800K.
    \item For certain GBs, namely the $\Sigma 3(211)$ and the $\Sigma 9(221)$ GBs, there are \emph{discontinuous} changes in the Zr distribution at certain Zr concentrations
    due to first-order complexion transitions.
    \item The distribution of segregated Zr at GBs is GB-specific and non-trivial. E.g. in some GBs the distribution of Zr atoms 
    breaks the mirror symmetry which is inherent in the GB for pure ceria.
    \item Doping ceria with Zr acts to reduce the formation enthalpies of most GBs, and can significantly affect the stabilities of different GBs
    relative to each other. E.g. at the miscibility gap at 1000K--1800K the $\Sigma 5(210)$ and $\Sigma 5(310)$
    GBs have formation enthalpies comparable to the $\Sigma 3(111)$ GB -- which is the most stable in pure ceria. 
    \item Changes in GB stability with Zr concentration have implications for the microstructure of ceria-zirconia. E.g. we predict that doping ceria with Zr at 1000K--1800K should result 
    in a more diverse range of GBs being exhibited by the Ce-rich phase.
\end{enumerate}

These predictions reflect the ability of SGCMC to capture important atomic-scale features of real grain boundaries, 
and demonstrate that the method should prove useful for studying GBs in metal oxide solid solutions. For instance, 
one possible application of SGCMC is to generate equilibrium GB structures which are later subject to further study using other theoretical methods, e.g. 
DFT to calculate the electronic structure of the GBs, or molecular dynamics to study diffusion at the GBs.

\section{Methods}
A detailed description of our methodology \cite{Kofke1988,Landau2009,Frenkel2002,Dubbeldam2013,Smith1995,Fitzgerald1999,Ferrenberg1988,Wilding2001,Errington2003,Bruce2003,Iba2001,Navrotsky2005,Watson1996,Symington2019,Symington2020,Williams2015,Purton2013,Brukhno2019,dlmontepython,Cabanas2001,Lee2008,ovito} and the interatomic potential we used 
in our calculations \cite{Sayle2013,Pedone2006}, as well as further computational details can be found in the supplementary 
information. The workhorse for our calculations was the open-source Monte Carlo simulation program DL\_MONTE \cite{Purton2013,Brukhno2019}.

\section{Data Availability}
Data pertaining to this work, including input and output files for our DL\_MONTE simulations, and scripts to perform data analysis and generate 
figures, can be found at \url{http://doi.org/10.5281/zenodo.8414910}.

\section{Acknowledgements}
TLU acknowledges support from the Engineering and Physical Sciences Research Council (EPSRC), grant
numbers EP/P007821/1 and EP/R023603/1. Via our membership of the UK's HEC Materials Chemistry Consortium, which is 
funded by EPSRC (EP/R029431/1 and EP/X035859/1), this work used the ARCHER2 UK National Supercomputing Service 
(\url{http://www.archer2.ac.uk}). This work also used the University of Bath's HPC facilities 
(\url{http://doi.org/10.15125/b6cd-s854}). MM and SV thank the University of Huddersfield for access to the Orion computing facility 
and the Violeta HPC.

\typeout{}

\bibliography{references.bib}


\appendix

\clearpage

\title{Supplementary Material for `Grain boundary segregation and phase separation in ceria-zirconia from atomistic simulation'}

\maketitle

\counterwithin{figure}{section}
\counterwithin{table}{section}

\section{Methodology}\label{sec:method}

\subsection{Semi-grand canonical Monte Carlo}
Semi-grand canonical Monte Carlo (SGCMC)\cite{Kofke1988,Frenkel2002,Landau2009} entails using the Metropolis Monte Carlo method to sample microstates of a 
system drawn from a semi-grand thermodynamical ensemble.
Here we consider an isothermal-isobaric semi-grand ensemble in which the prescribed thermodynamic parameters are: the total number of cations, 
$N\equiv(N_{\text{Ce}}+N_{\text{Zr}})$, where $N_{\text{Ce(Zr)}}$ denotes the number of Ce(Zr) atoms in the system; the pressure, $P$;
the temperature, $T$; and the chemical potential \emph{difference} between the two cation species, $\Delta\mu\equiv(\mu_{\text{Zr}}-\mu_{\text{Ce}})$.
Moreover, in this work we only consider the case $P=0$. 
Note that here we ignore complications such as O vacancies; the number of O atoms is fixed by $N$ to be $N_{\text{O}}=2N$, and the total number of
atoms in the system is $\mathcal{N}=(N+N_{\text{O}})=3N$. 
We also assume that the system we consider is orthorhombic, with dimensions $L_x$, $L_y$ and $L_z$ along the Cartesian axes.
Thus a microstate of the system constitutes a specification of $L_x$, $L_y$, $L_z$, the positions of the atoms, and the species of the atoms.
In this ensemble the probability of the system being in microstate $i$ is given by
\begin{equation}
p_i=\frac{1}{\Psi}\exp\Bigl[-\beta\bigl(E_i+PV_i-\Delta\mu N_{\text{Zr},i}\bigr)\Bigr],
\end{equation}
where $E_i$ is the energy of the microstate, $V_i$ is its volume, $N_{\text{Zr},i}$ is the number of Zr atoms for the microstate, 
$\beta\equiv 1/(k_BT)$ is the thermodynamic beta, 
$k_B$ is the Boltzmann constant, and $\Psi$ is the partition function for the ensemble. 

We use the following three types of Monte Carlo move to sample the phase space. Firstly, we use
\emph{translation moves} in which a Ce, Zr or O atom is chosen at random and its position is translated by a random amount.
Secondly, we use \emph{volume moves} in which the volume of the system is expanded or contracted by a random amount \emph{along a 
randomly chosen Cartesian dimension}. In other words, we employ volume moves which change one of $L_x$, $L_y$ or $L_z$.
Finally, we use \emph{transformation moves} in which a Ce atom is selected at random, and its species is transformed to Zr,
or vice versa. Note that a transformation move changes the number of Ce and Zr atoms in the system, $N_{\text{Ce}}$ and $N_{\text{Zr}}$, while preserving the
total number of cations $N$. Note also that the interatomic potentials we use in our calculations (which we describe later in Section~\ref{sec:model}) 
always use the same partial charge for Ce and Zr atoms, and hence these transformation moves maintain charge neutrality of the system.
Further details regarding SGCMC can be found elsewhere \cite{Frenkel2002,Dubbeldam2013}. 

\subsubsection{Relating $\Delta\mu$ to composition $c$}
Arguably one drawback of the semi-grand ensemble is that $\Delta\mu$ cannot easily be related to experimental
conditions. Instead of $\Delta\mu$, it is the bulk concentrations of Ce and Zr, i.e. $c=N_{\text{Zr}}/N$ and
$(1-c)=N_{\text{Ce}}/N$, which are imposed in experimental studies of ceria-zirconia; and we are interested in using
simulation to determine how a target physical quantity, $X$, (e.g. density) varies with $c$ and $T$ -- as opposed to $\Delta\mu$ and $T$. 
This issue can be overcome by noting that $c$ is completely determined by $T$ and $\Delta\mu$, and hence tuning $\Delta\mu$ (while keeping $T$ 
fixed) amounts to tuning $c$. The $c$ corresponding to a given $\Delta\mu$ can be obtained straightforwardly from a SGMC simulation of the bulk 
via $c=\langle N_{\text{Zr}}\rangle/N$, where $\langle N_{\text{Zr}}\rangle$ is the mean number of Zr cations observed in the 
$N$-cation bulk system during the simulation. Hence any physical quantity $X$ can be obtained for a given $c$ by: 1) 
performing exploratory simulations to determine the $\Delta\mu$ which yields the $c$ of interest at the considered $T$; 2) performing a SGCMC 
simulation at the $\Delta\mu$ corresponding to that $c$, and then retrieving $X$ from the simulation data.

One of the aims of this work is to study the properties of GBs at prescribed $T$ and bulk composition
$c$. With the above in mind, this entailed first performing, for each considered $T$, a series of bulk SGCMC simulations to determine the 
$\Delta\mu$ which corresponded to the values of $c$ of interest. These $\Delta\mu$ were then used in SGCMC simulations of the GB 
systems, yielding GB properties which corresponded to known bulk compositions.

\subsection{Calculating the miscibility gap using free energy methods}\label{sec:miscibility_gap_method}
At certain $T$ we expect ceria-zirconia for our choice of model (described later in Section~\ref{sec:model}) to exhibit phase
separation into two phases, a Ce-rich phase and a Zr-rich phase, for $c$ within a range of global compositions $c_1<c<c_2$. Such global compositions for all
$T$ constitutes the miscibility gap.
For global compositions within the miscibility gap, the Ce- and Zr-rich phases take the compositions $c_1$ and $c_2$, respectively, for the 
considered $T$. One of our aims is to calculate the miscibility gap, i.e. calculate $c_1$ and $c_2$ for a range of temperatures. 
Moreover, we emphasise that we are interested in the $c_1$ and $c_2$ which correspond to \emph{thermodynamic equilibrium}, as opposed to the limits of
metastability of a solid solution to phase separation. In other words we are interested in the $c_1$ and $c_2$ which constitute the \emph{binodal}
compositions for the miscibility gap, as opposed to the spinodal compositions.

At a temperature $T$ exhibiting phase separation the composition $c$ as a function of $\Delta\mu$ exhibits a 
discontinuity at a certain chemical potential difference $\Delta\mu_{\text{co}}$, jumping from composition $c_1$ to composition $c_2$. 
This reflects the fact that phase separation is a manifestation of a first-order phase transition from the Ce-rich phase to the Zr-rich phase
at $\Delta\mu_{\text{co}}$.
With this in mind, one could in principle determine $c_1$ and $c_2$ from inspection of a plot of $c$ versus $\Delta\mu$ at $T$ obtained using SGCMC. 
However, in practice hysteresis effects can prevent $c_1$ and $c_2$ from being pinpointed in tractable simulation timescales. 
The issue is that in the vicinity of a first order phase transition there is a free energy barrier which inhibits movement between the 
Ce-rich and Zr-rich phases in reasonable timescales; the system `gets stuck' in the phase in which the simulation was 
initialised. 

Fortunately, methods exist which alleviate this problem, enabling $c_1$ and $c_2$ to be pinpointed. 
Here we use two methods, namely transition-matrix MC (TMMC)\cite{Smith1995,Fitzgerald1999}
and histogram reweighting (HR) \cite{Ferrenberg1988,Landau2009}, 
to efficiently determine $\Delta\mu_{\text{co}}$, $c_1$ and $c_2$ at various $T$ essentially exactly. We elaborate on this below.
Our use of TMMC and HR closely resembles previous applications of these methods to the liquid-gas transition in fluids. 
Hence below we provide only a concise description of these methods, focusing on the aspects which differ in the semi-grand 
ensemble. Further details regarding TMMC and HR in this context can be found elsewhere \cite{Wilding2001,Errington2003}.

\subsubsection{Transition-matrix Monte Carlo}
TMMC \cite{Smith1995,Fitzgerald1999} belongs to a class of methods \cite{Bruce2003,Iba2001} in which a bias is added to the Hamiltonian which serves to
eliminate the free energy barrier separating the phases. The bias is a function of some order parameter appropriate to the problem at hand.
In our case the order parameter is $N_{\text{Zr}}$: 
low values of $N_{\text{Zr}}$ correspond to a Ce-rich phase, high values correspond to a Zr-rich phase, and thus $N_{\text{Zr}}$ parametrises a pathway
linking the two phases. We use TMMC to determine the free energy associated with $N_{\text{Zr}}$ at certain $T$ and $\Delta\mu$:
\begin{equation}\label{eqn:pNCe}
    F(N_{\text{Zr}})\equiv-\frac{1}{\beta}\ln{p(N_{\text{Zr}})},
\end{equation}
where
\begin{equation}
    p(N_{\text{Zr}})=\frac{1}{\Psi}\sum_i\delta_{N_{\text{Zr}}N_{\text{Zr},i}}\exp\Bigl[-\beta\bigl(E_i+PV_i-\Delta\mu N_{\text{Zr},i}\bigr)\Bigr]
\end{equation}
is the probability associated with $N_{\text{Zr}}$, the summation in the above equation is over all microstates $i$,
and $\delta_{N_{\text{Zr}}N_{\text{Zr},i}}$ is a Kronecker delta.
To elaborate, in our TMMC simulations microstates are sampled with probabilities
\begin{equation}
\tilde{p}_i=\frac{1}{\tilde{\Psi}}\exp\Bigl[-\beta\bigl(E_i+PV_i-\Delta\mu N_{\text{Zr},i}\bigr)+\eta(N_{\text{Zr},i})\Bigr],
\end{equation}
where $\eta(N_{\text{Zr}})$ is the \emph{bias} applied to microstates with order parameter $N_{\text{Zr}}$, and
$\tilde{\Psi}$ is the partition function of the corresponding biased ensemble. The key aspect of TMMC is that
$\eta(N_{\text{Zr}})$ is calculated on the fly during the simulation, with $\eta(N_\text{Zr})$ being calculated from statistics pertaining to the
transitions between different values of $N_{\text{Zr}}$. Moreover, $\eta(N_\text{Zr})$ is evolved so that it converges upon a function which yields 
uniform sampling over all values of $N_{\text{Zr}}$. Once $\eta(N_\text{Zr})$ converges upon this target function it can be shown that the 
free energy associated with order parameter $N_{\text{Zr}}$, which we denote as $F(N_{\text{Zr}})$, is given by
\begin{equation}
\beta F(N_{\text{Zr}})=\eta(N_{\text{Zr}})+C,
\end{equation}
where $C$ is an arbitrary constant.
Therefore, upon convergence of $\eta(N_{\text{Zr}})$, $F(N_{\text{Zr}})$ can be obtained from $\eta(N_{\text{Zr}})$ by using the above equation.
Moreover, from Eqn.~\ref{eqn:pNCe} $p(N_{\text{Zr}})$ follows trivially.

\subsubsection{Calculating $c_1$ and $c_2$}
Much information about the system can be obtained from $F(N_{\text{Zr}})$ or, because it contains the same information, $p(N_{\text{Zr}})$. 
For $T$ where the system exhibits a miscibility gap, at $\Delta\mu$ close to $\Delta\mu_{\text{co}}$, $F(N_{\text{Zr}})$ exhibits two local minima, one at a low value
of $N_{\text{Zr}}$ corresponding to the Ce-rich phase and one at a high value of $N_{\text{Zr}}$ corresponding to the Zr-rich phase. The free energy difference
between the Ce-rich and Zr-rich phases, $\Delta F\equiv (F_{\text{Zr-rich}}-F_{\text{Ce-rich}})$, can be evaluated from $p(N_{\text{Zr}})$ using
\begin{equation}\label{eqn:DeltaF}
    \beta\Delta F \equiv -\ln\Biggl[ \frac{\sum_{N_{\text{Zr}}>N_{\text{thresh}}}p(N_{\text{Zr}})}{\sum_{N_{\text{Zr}}<N_{\text{thresh}}}p(N_{\text{Zr}})} \Biggr],
\end{equation}
where $N_{\text{thresh}}$ is a value of $N_{\text{Zr}}$ which defines the Ce-rich and Zr-rich phases: microstates with $N_{\text{Zr}}>N_{\text{thresh}}$
constitute the Zr-rich phase.
Moreover, the compositions of the Ce-rich and Zr-rich phases are given by
\begin{equation}\label{eqn:equil_composition_Zr_rich}
   \langle c\rangle_{\text{Zr-rich}} = \frac{1}{N}\sum_{N_{\text{Zr}}>N_{\text{thresh}}}p(N_{\text{Zr}})N_{\text{Zr}}
\end{equation}
and
\begin{equation}\label{eqn:equil_composition_Ce_rich}
   \langle c\rangle_{\text{Ce-rich}} =  \frac{1}{N}\sum_{N_{\text{Zr}}<N_{\text{thresh}}}p(N_{\text{Zr}})N_{\text{Zr}}
\end{equation}
respectively.
By definition $c_1$ and $c_2$ are the compositions of the Ce-rich and Zr-rich phases at $\Delta\mu_{\text{co}}$, which corresponds to when $\Delta F=0$. 
Hence $c_1$ and $c_2$, can be obtained by applying Eqns.~\ref{eqn:equil_composition_Zr_rich} and \ref{eqn:equil_composition_Ce_rich} at $\Delta\mu_{\text{co}}$. 

However, there is the question of how to locate $\Delta\mu_{\text{co}}$ in order to evaluate these quantities. In principle one could perform SGCMC simulations 
at a series of $\Delta\mu$ in order to locate $\Delta\mu_{\text{co}}$,
at each $\Delta\mu$ calculating $p(N_{\text{Zr}})$, and then $\Delta F$ via Eqn.~\ref{eqn:DeltaF} for each $\Delta\mu$ in a search for the $\Delta\mu$ which
yields $\Delta F=0$. However, a more
elegant approach is to use \emph{histogram reweighting} (HR)\cite{Landau2009,Ferrenberg1988} to pinpoint $\Delta\mu_{\text{co}}$ using data from a 
single simulation at a certain $\Delta\mu$ `very close' to $\Delta\mu_{\text{co}}$. We used this approach here. To elaborate, 
HR \cite{Landau2009,Ferrenberg1988} is a method for
using data obtained from simulation at one set of thermodynamic parameters to calculate physical properties at a different set of thermodynamic parameters
not directly explored in the simulation.
In our case we used HR to calculate $F(N_{\text{Zr}})$ at values of $\Delta\mu$ close to that used in the TMMC simulation. The relevant equation is
as follows:
\begin{equation}\label{eqn:HR}
    \beta F(N_{\text{Zr}};\Delta\mu')=\beta F(N_{\text{Zr}};\Delta\mu_{\text{sim}})-\beta(\Delta\mu'-\Delta\mu_{\text{sim}})N_{\text{Zr}},
\end{equation}
where $F(N_{\text{Zr}};\Delta\mu)$ denotes the function $F(N_{\text{Zr}})$ at $\Delta\mu$, $\Delta\mu_{\text{sim}}$ is the value of $\Delta\mu$ which the
simulation was performed at, and $\Delta\mu'$ is the value of $\Delta\mu$ we are extrapolating to. 

Thus our procedure to pinpoint $\Delta\mu_{\text{co}}$ at a given temperature $T$ was as follows. 
First, we performed a TMMC simulation at a value of $\Delta\mu_{\text{sim}}$ close to $\Delta\mu_{\text{co}}$,
obtaining $F(N_{\text{Zr}})$ from the converged $\eta(N_{\text{Zr}})$ as described above.
(Exploratory simulations were used to find a value of $\Delta\mu_{\text{sim}}$ sufficiently close to $\Delta\mu_{\text{co}}$).
We then used HR to determine $F(N_{\text{Zr}})$, and then $\Delta F$, at a range of $\Delta\mu'$ near $\Delta\mu_{\text{sim}}$, in a search for phase
coexistence. By definition coexistence occurs when $\Delta F=0$; we used an optimisation procedure to pinpoint
$\Delta\mu_{\text{co}}$ using this criterion. 
Once $\Delta\mu_{\text{co}}$ has been determined, the binodal compositions $c_1$ and $c_2$ were obtained as described as described above.

\subsection{Simulations at fixed composition}
In the semi-grand ensemble $\Delta\mu$ is imposed, and $c$ exhibits fluctuations. We employed this ensemble for the majority of our
calculations. However, we also performed a small number of calculations in an $NPT$ ensemble where $c$ was fixed. These calculations
served to facilitate comparison with previous studies. To elaborate, we considered fluorite-like bulk systems at 300K with compositions ranging from
$c=0$ to $c=1$, with the initial positions of Ce and Zr atoms distributed randomly on the cation sublattice. 
Moreover, only translational and volume Monte Carlo moves were used to evolve the system. 
These calculations were used to determine the lattice parameter and enthalpy of mixing of the solid solution versus $c$. 

\subsubsection{Enthalpy of mixing}
The enthalpy of mixing we consider here is that associated with creation of the cubic (fluorite) ceria-zirconia solid solution with
composition $c$ from cubic ceria and \emph{monoclinic} zirconia -- which is the stable polymorph of zirconia at 300K. This quantity is defined as
\begin{align}\label{eqn:H_form}
\Delta H_{\text{mix}}(c) =  & H[\text{c-Ce}_{1-c}\text{Zr}_c\text{O}_2] \nonumber \\
                        -& \Bigl\lbrace(1-c)H[\text{c-Ce}\text{O}_2]+cH[\text{m-Zr}\text{O}_2]\Bigr\rbrace, 
\end{align}
where $\text{c-Ce}\text{O}_2$ and $\text{c-Ce}_{1-c}\text{Zr}_c\text{O}_2$ denote cubic
ceria and ceria-zirconia, respectively; $\text{m-Zr}\text{O}_2$ denotes monoclinic
zirconia; and all quantities in this equation are enthalpies per $MO_2$ unit, where $M$ is a cation. 
Defining 
\begin{equation}
\Delta H_{\text{m}\to\text{c}}= H[\text{c-Zr}\text{O}_2]-H[\text{m-Zr}\text{O}_2]
\end{equation}
as the enthalpy change associated with transforming ZrO$_2$ from the monoclinic phase to the cubic phase, Eqn.~\ref{eqn:H_form}
can be written as
\begin{equation}
\Delta H_{\text{mix}}(c) = \Delta H_{\text{c-mix}}(c) +  c\Delta H_{\text{m}\to\text{c}},
\end{equation}
where
\begin{align}
\Delta H_{\text{c-mix}}(c)= & H[\text{c-Ce}_{1-c}\text{Zr}_c\text{O}_2] \nonumber \\
               -& \Bigl\lbrace (1-c)H[\text{c-Ce}\text{O}_2]+cH[\text{c-Zr}\text{O}_2]\Bigr\rbrace
\end{align}
is the enthalpy of mixing for a ceria-zirconia solid solution from cubic ceria and and cubic zirconia.
Using a similar approach to taken in other studies\cite{Lee2008,Grau-Crespo2011},
we obtained $\Delta H_{\text{mix}}(c)$ by combining $\Delta H_{\text{c-mix}}(c)$ obtained from simulations of the cubic phase with an experimental value of 
$\Delta H_{\text{m}\to\text{c}}=8.8$\,kJ/mol \cite{Navrotsky2005}.

\subsection{Grain boundary properties}
After calculating the aforementioned bulk properties of ceria-zirconia, we turned to studying
GBs, focusing on the Ce-rich phase at various temperatures. 
For each GB we performed SGCMC simulations for a range of $\Delta\mu$ up to $\Delta\mu_{\text{co}}$, or, equivalently, a range of bulk compositions 
$c$ up to the binodal composition for the Ce-rich phase, $c_1$.
To characterise the structure and stability of the GBs we used a number of metrics, which we will now describe.

\subsubsection{Quantifying segregation}
We were particularly interested in the overall quantity of Zr segregation to the GBs. We used the following metric for the
segregation per unit area of the interface:
\begin{equation}\label{eqn:seg_metric}
    \xi = \frac{1}{A}\Bigl[\langle N_{\text{Zr}}\rangle-Nc^{\text{bulk}}\Bigr],
\end{equation}
where $A$ is the cross-sectional area of the GB in the simulated GB system, $\langle N_{\text{Zr}}\rangle$ is the 
observed number of Zr atoms in the GB system at equilibrium, $N$ is the number of cation sites in the system, and $c^{\text{bulk}}$ is the \emph{bulk} Zr 
concentration corresponding to the $\Delta\mu$ under consideration. Note that $[\langle N_{\text{Zr}}\rangle-Nc^{\text{bulk}}]$ is the \emph{excess} 
number of Zr atoms in the GB system relative to a bulk system with the same number of cations, and hence $\xi$ is the excess number of 
Zr atoms at the GB per unit area of the GB due to segregation.
Note that this measure of segregation is convenient because it does not require one to define a region of space 
associated with the GB. 

As well as the overall amount of segregation to a GB, we are also interested in the specific
distribution of Zr atoms in space at the GB. To characterise the distribution of Zr \emph{across} the GB, 
i.e. the \emph{segregation profile}, we used 
\begin{equation}\label{eqn:seg_density}
    \rho_{\text{Zr}}(z)=\biggl\langle\sum_{j\in\text{Zr}}\delta(z_j-z)\biggr\rangle,
\end{equation}
where $\delta(z)$ here is a Dirac delta function, the summation is over all Zr atoms $j$, $z_j$ is the $z$ coordinate of atom $j$, $z$ is the
dimension perpendicular to the GB, and $\langle\dotsc\rangle$ denotes an ensemble average for the $\Delta\mu$ and $T$
under consideration.

\subsubsection{Quantifying stability}
Stability in our isothermal-isobaric semi-grand ensemble is determined by the following thermodynamic potential, which is obtained
via a Legendre transformation of the energy in the usual manner:
\begin{equation}
\Omega=E+PV-TS-\Delta\mu N_{\text{Zr}}.
\end{equation}
Hence the appropriate measure for thermodynamic stability of a GB is the change in $\Omega$ upon formation
of unit area of the GB, i.e.
\begin{equation}\label{eqn:Omega_GB}
    \Delta\Omega_{\text{form}}=(\Omega_{\text{GB}}-\Omega_{\text{bulk}})/A,
\end{equation}
where $\Omega_{\text{GB}}$ is the value of $\Omega$ for a system exhibiting area $A$ of the GB at the considered $T$ and composition, and 
$\Omega_{\text{bulk}}$ is the value of $\Omega$ for a bulk system (exhibiting no GB) at the same conditions and of the same
size (i.e. contains the same number of cations and O atoms) as the GB system.
In principle $\Delta\Omega_{\text{form}}$ could be calculated using simulation. 
However, in practice this is difficult since it would require evaluation of a
difference in entropy between the two systems, $\Delta S_{\text{form}}\equiv (S_{\text{GB}}-S_{\text{bulk}})/A$. We do not do this here. 

Instead we evaluate what we refer to here as the \emph{GB formation enthalpy}, $\Delta J_{\text{form}}$, which is 
$\Delta\Omega_{\text{form}}$ ignoring the contribution from $\Delta S_{\text{form}}$. To elaborate, it is convenient to define a quantity
\begin{equation}
J \equiv E+PV-\Delta\mu N_{\text{Zr}},
\end{equation}
which, on account of the fact that it does not include a $TS$ term, is relatively easy to calculate for a given system using simulation -- unlike
$\Omega$. $J$ in the isothermal-isobaric semi-grand ensemble is analogous to $E$ in the canonical ensemble, or the 
enthalpy $H$ in the isothermal-isobaric ensemble. 
Thus the GB formation enthalpy is defined as
\begin{equation}\label{eqn:GB_formation_enthalpy}
    \Delta J_{\text{form}}=(J_{\text{GB}}-J_{\text{bulk}})/A
\end{equation}
where, similarly to Eqn.~\ref{eqn:Omega_GB}, $J_{\text{GB}}$ pertains to a system with area $A$ of the GB under consideration and
$J_{\text{bulk}}$ pertains to a bulk system of the same size.
Note that in practice to calculate $\Delta J_{\text{form}}$ we did not use pairs of GB and bulk systems of equal size. Rather, to calculate
$\Delta J_{\text{form}}$ we rescaled the value of $J_{\text{bulk}}$ obtained from the bulk simulation so that it corresponded to the system size (i.e.
number of cations $N$) used to evaluate $J_{\text{GB}}$.

\subsection{System details}
\subsubsection{Bulk systems}
The systems used in our bulk simulations were comprised of 1500 atoms, i.e. 500 cations (Ce or Zr) and 1000 O atoms, and were 
formed by tiling 5 12-atom fluorite unit cells in each of the Cartesian directions.
For our fixed-composition bulk simulations the species of cations were assigned at random so that the composition $c$ reflected that of
interest. By contrast for our SGCMC simulations, all cations were initially assigned to be Ce. Moreover, recall that in all of our
simulations $P=0$ and the volume moves were such that system can expand and contract independently along each Cartesian dimension. 
This allows the system freedom to equilibrate to either the cubic or tetragonal fluorite-like structure depending on the thermodynamic conditions.

\subsubsection{Grain boundary systems}
\begin{figure*}
	\includegraphics[width=2.0\columnwidth]{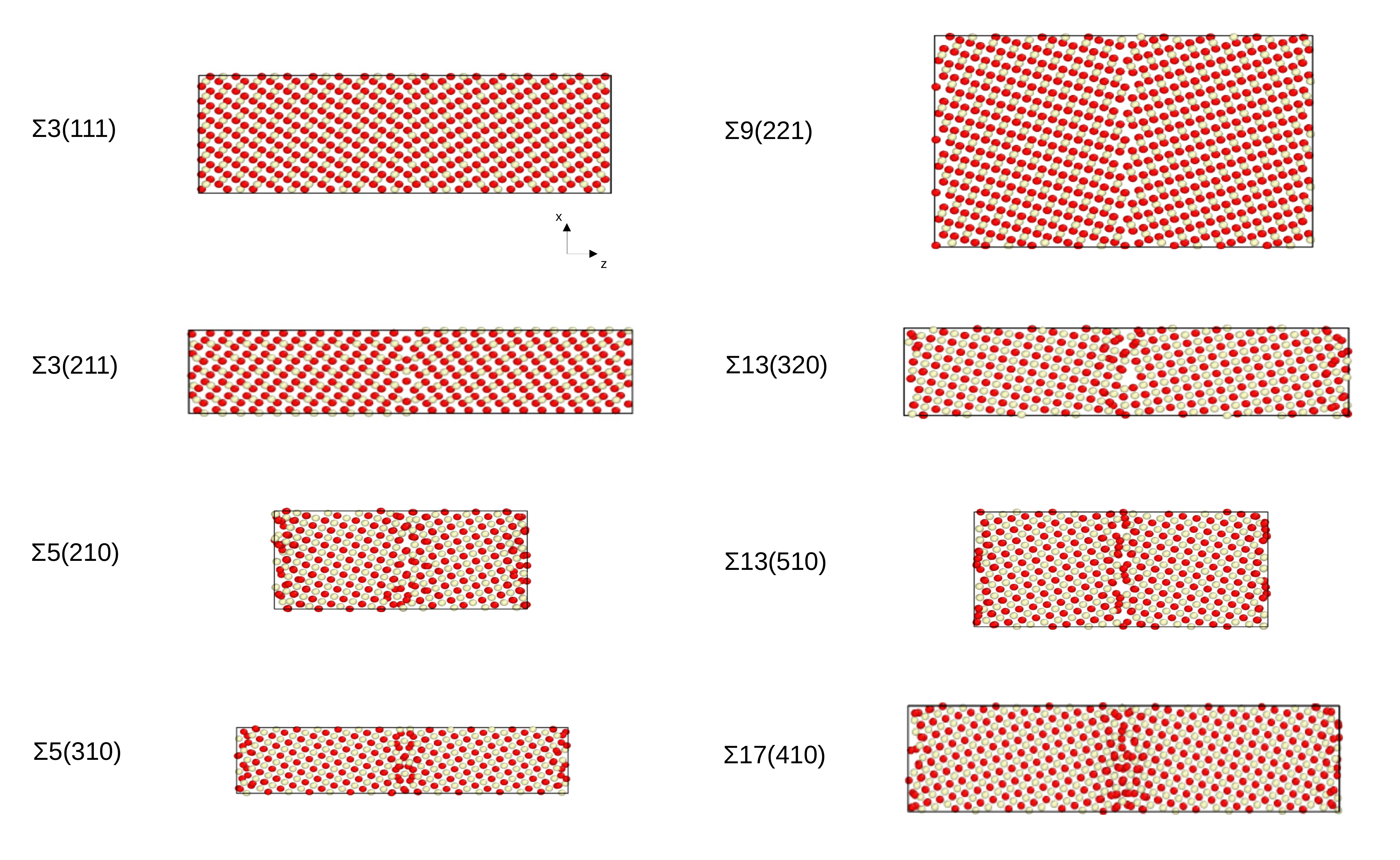}
	\caption{The initial structures, generated as described in the text, used in our SGCMC simulations of GBs. Each structure
 corresponds to a different GB type, whose name is given to the left of the structure. Ce atoms are coloured light yellow while O 
 atoms are coloured red. All structures are shown in the x/z plane, where the z direction (which is horizontal to the page) is defined to 
 be perpendicular to the GB plane.} 
	\label{fig:GB_supercells}
\end{figure*}

The GBs we considered were the $\Sigma 3(111)$, $\Sigma 3(211)$, $\Sigma 5(210)$, $\Sigma 5(310)$, $\Sigma 9(221)$ $\Sigma 13(320)$, 
$\Sigma 13(510)$ and $\Sigma 17(410)$ twin GBs with a fluorite underlying crystal. 
The GB structures we used as initial configurations in our simulations corresponded to pure ceria. These were 
generated using the METADISE code \cite{Watson1996} and the procedure described in previous studies 
\cite{Watson1996,Symington2019,Symington2020,Williams2015}, 
followed by annealing at 10K and zero pressure using translational and volume Monte Carlo moves
to effectively minimise the energy of the structures.
The structures were comprised of between $\approx$2000--5000 atoms, and
were designed so that: 1) the distance between the two GBs in the structure was at least 50\AA; 2) the 
structure was orthorhombic in shape; 3) the structure's dimensions were at least $2R_C$ in extent in each Cartesian direction
at all thermodynamic conditions we considered, where $R_C$ is the cut-off for the short-range interactions in our choice of 
model. 
Table~\ref{tab:GB_geometry} provides the number of atoms in each of the structures, as well as their Cartesian dimensions.
Images of the structures are provided in Fig.~\ref{fig:GB_supercells}.

\begin{table}
    \begin{ruledtabular}
        \begin{tabular}{cccccc}
             GB type & $\mathcal{N}$ & $L_x$ (\AA) & $L_y$ (\AA) & $L_z$ (\AA) \\
             \hline
%
             $\Sigma 3(111)$ & 3456    &  26.50   &  22.95   &  75.03   \\
             $\Sigma 3(211)$ & 2592    &  18.73   &  22.95   &  80.55   \\
             $\Sigma 5(210)$ & 1920    &  24.20   &  21.64   &  50.40   \\
             $\Sigma 5(310)$ & 1920    &  17.06   &  21.65   &  69.78   \\
             $\Sigma 9(221)$ & 5184    &  45.91   &  22.93   &  66.21   \\
             $\Sigma 13(320)$& 2496    &  19.52   &  21.65   &  80.27   \\
             $\Sigma 13(510)$& 2496    &  27.58   &  21.66   &  57.02    \\
             $\Sigma 17(410)$& 3264    &  22.31   &  21.65   &  91.21   \\ 
        \end{tabular}
    \end{ruledtabular}
    \caption{Geometrical information regarding the structures used as initial configurations in our GB simulations. 
    For each type of GB considered we provide the total number of atoms in the structure ($\mathcal{N}$) and its Cartesian 
    dimensions ($L_x$, $L_y$ and $L_z$). The $z$ direction is defined here to be perpendicular to the GB planes.}
    \label{tab:GB_geometry}
\end{table}

\subsection{Interatomic potential}\label{sec:model}
In our calculations the interaction energy a pair of atoms belonging to elements $m$ and $n$ is given by
\begin{equation}\label{eqn:potential}
    U_{mn}(r)=k_e\frac{q_mq_n}{r}+D_{mn}\Bigl\lbrace\bigl[1-e^{-a_{mn}(r-r_{0mn})}\bigr]^2-1\Bigr\rbrace+\frac{C_{mn}}{r^{12}},
\end{equation}
where $r$ is the separation between the two atoms in question; $k_e=1/(4\pi\epsilon_0)$ is the Coulomb constant;
and the remaining variables $q_m$, $q_n$, $D_{mn}$, $a_{mn}$, $r_{0mn}$ and $C_{mn}$ parameterise the interaction between
the elements. The first term in the above equation models the long-range component of the interactions between $m$ and $n$ atoms. In
this term, $q_m$ is the partial charge associated with element $m$, and similarly for $q_n$. In our calculations we used $q_{\text{Ce}}=2.4e$, 
$q_{\text{Zr}}=2.4e$ and $q_{\text{O}}=-1.2e$, where $e$ denotes the proton charge. The second and third terms in the above equation
constitute the short-range component of the interaction: a Morse potential and a $1/r^{12}$ hard-core repulsion, respectively. 
Our choice of $D_{mn}$, $a_{mn}$, $r_{0mn}$ and $C_{mn}$
for each pair of elements in our system is given in Table~\ref{tab:potential}. However, note that
parameters for short-range interactions between pairs of cation elements, namely Ce-Ce, Ce-Zr and Zr-Zr, are not listed in this 
table. 
This reflects the fact that there are no such interactions in our model: in effect $D_{mn}=0$ and $C_{mn}=0$ for such pairs.
The studies from which we obtained our choice of parameters are provided in the table.
Note also that we used a cut-off of $R_C=8.5$\AA\ in our calculations, and did not apply long-range corrections to the $1/r^{12}$
term. 

\begin{table}
    \begin{ruledtabular}
        \begin{tabular}{cccccc}
            Pair & $D_{mn}$ (eV) & $a_{mn}$ (\AA$^{-1}$) & $r_{0mn}$ (\AA) & $C_{mn}$ (eV\AA$^{12}$) & Ref. \\
            \hline
            Ce--O            &      0.098352        &        1.848592             &      2.930147          &     1.0   &  \onlinecite{Sayle2013}  \\
            O--O             &      0.206237        &        2.479675             &      2.436997          &     1.0   &  \onlinecite{Sayle2013}  \\
            Zr--O            &      0.041730        &        1.886824             &      3.189367          &     22.0  &  \onlinecite{Pedone2006} \\

        \end{tabular}
    \end{ruledtabular}
    \caption{Parameters for the short-range component of the interatomic potential (Eqn.~\ref{eqn:potential}) used in our calculations for 
    various pairs of elements. Parameters for Ce-Ce, Ce-Zr and Zr-Zr are not listed because the short-range component to the interaction these pairs of elements is set to zero.}
    \label{tab:potential}
\end{table}

\subsection{Further simulation details}
Our SGCMC simulations employed Monte Carlo moves in the ratio translation:transformation:volume=1000:600:3;
while our fixed-composition simulations used the ratio translation:volume=1000:1. The sizes of translational and volume moves were chosen to yield approximately
37\% and 50\% success rates, respectively, and move sizes were fixed for the duration of a simulation.
Our free energy SGCMC simulations were performed for $\approx 3\times 10^9$ Monte Carlo moves, with updates of the bias function performed every
8000 moves.
Our fixed-composition bulk (to calculate lattice parameters and enthalpies of mixing versus Zr composition), 
and SGCMC GB simulations were performed for $1.5\times 10^7$ and $>8\times10^8$ moves, respectively. Moreover, to calculate 
$c^{\text{bulk}}$ in Eqn.~\ref{eqn:seg_metric}, we employed bulk SGCMC simulations of length $\approx 2\times 10^8$ moves at the 
same $\Delta\mu$ as we considered in the GB simulations. 
We applied block averaging to data, obtained after the initial equilibration period, to calculate 
uncertainties for physical quantities of interest, e.g. enthalpy, $\xi$ (Eqn.~\ref{eqn:seg_metric}).
All of our Monte Carlo calculations were performed using the general-purpose Monte Carlo simulation program DL\_MONTE \cite{Purton2013,Brukhno2019},
and the Python package dlmontepython \cite{dlmontepython} was used to apply histogram reweighting (Eqn.~\ref{eqn:HR}) to the results of the free energy calculations. 
DL\_MONTE employs an Ewald summation to calculate electrostatic energies. While it is conventional to use a fixed cut-off in reciprocal space to determine the
set of reciprocal lattice vectors used in the Ewald summation, we instead used a fixed set of vectors throughout each simulation. 
We used Ovito to visualise configurations output by our simulations, and generate the figures of configurations in this 
work \cite{ovito}.
Further details can be found in the data accompanying this work.

\section{Further results}

\subsection{Model validation: Lattice parameter and enthalpy of mixing}
Before undertaking SGCMC calculations we first
performed various calculations at fixed $c$ to assess the accuracy of our chosen model. To elaborate, 
we calculated the bulk lattice parameters and enthalpies of mixing (Eqn.~\ref{eqn:H_form}) at 300K
for various compositions $c$ of fluorite ceria-zirconia, and compared the results to experiment \cite{Cabanas2001,Lee2008} and
DFT-based theoretical calculations \cite{Grau-Crespo2011}. 
These calculations utilised bulk systems with a fixed composition of Ce and Zr atoms distributed randomly on the cation sublattice;
further details of the calculations are provided in Section~\ref{sec:method}.

\begin{figure}
	\includegraphics[width=1.0\columnwidth]{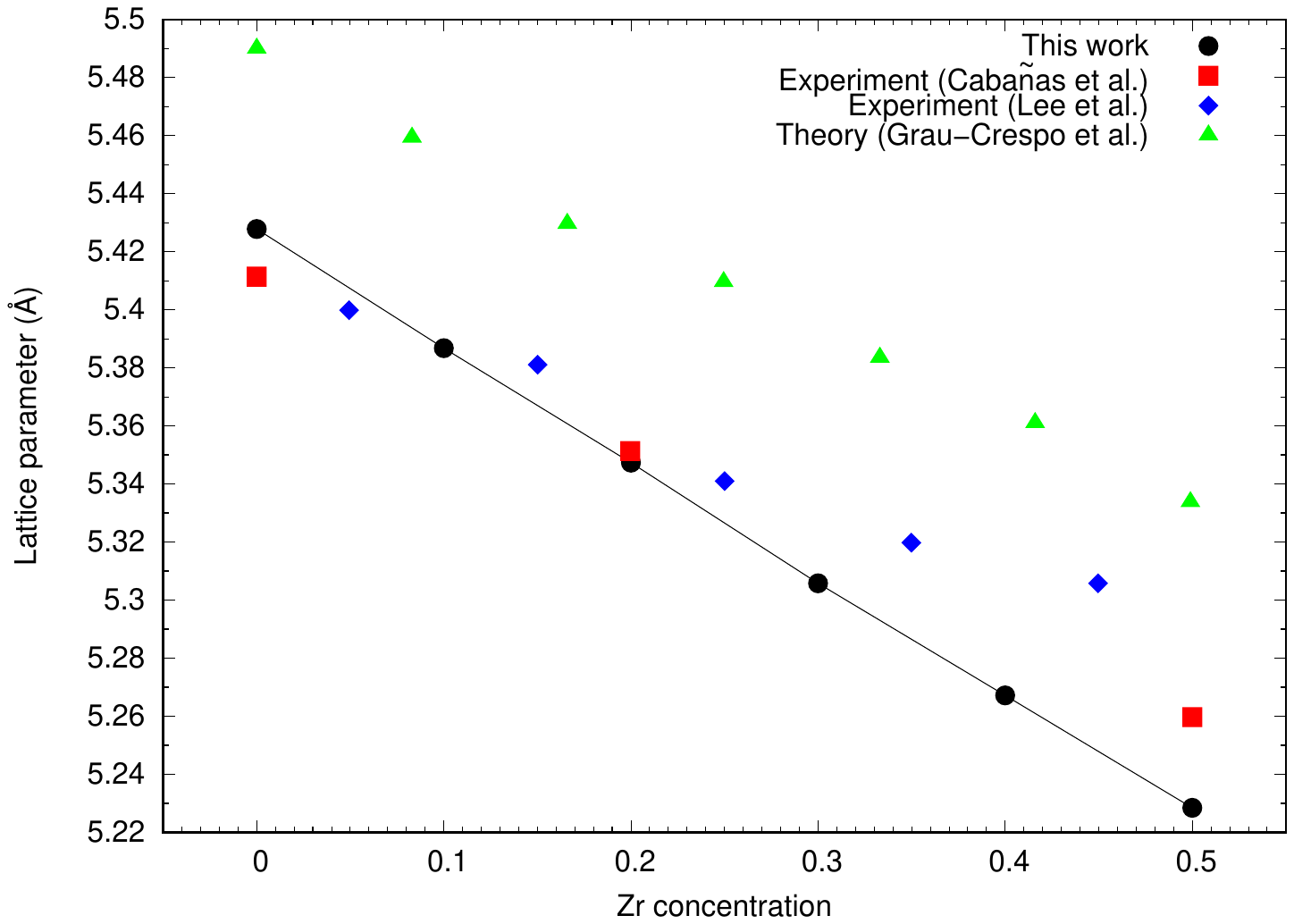}
	\caption{Lattice parameter versus Zr concentration for bulk fluorite ceria-zirconia 
	obtained from this work compared against previous experimental (Cabanas et al.\cite{Cabanas2001}
	and Lee et al. \cite{Lee2008}) and theoretical (Grau-Crespo et. al \cite{Grau-Crespo2011}) studies. The uncertainty in the results for this
        work are smaller than
	the size of the data points.} 
	\label{fig:lattice_parameter}
\end{figure}

\begin{figure}
	\includegraphics[width=1.0\columnwidth]{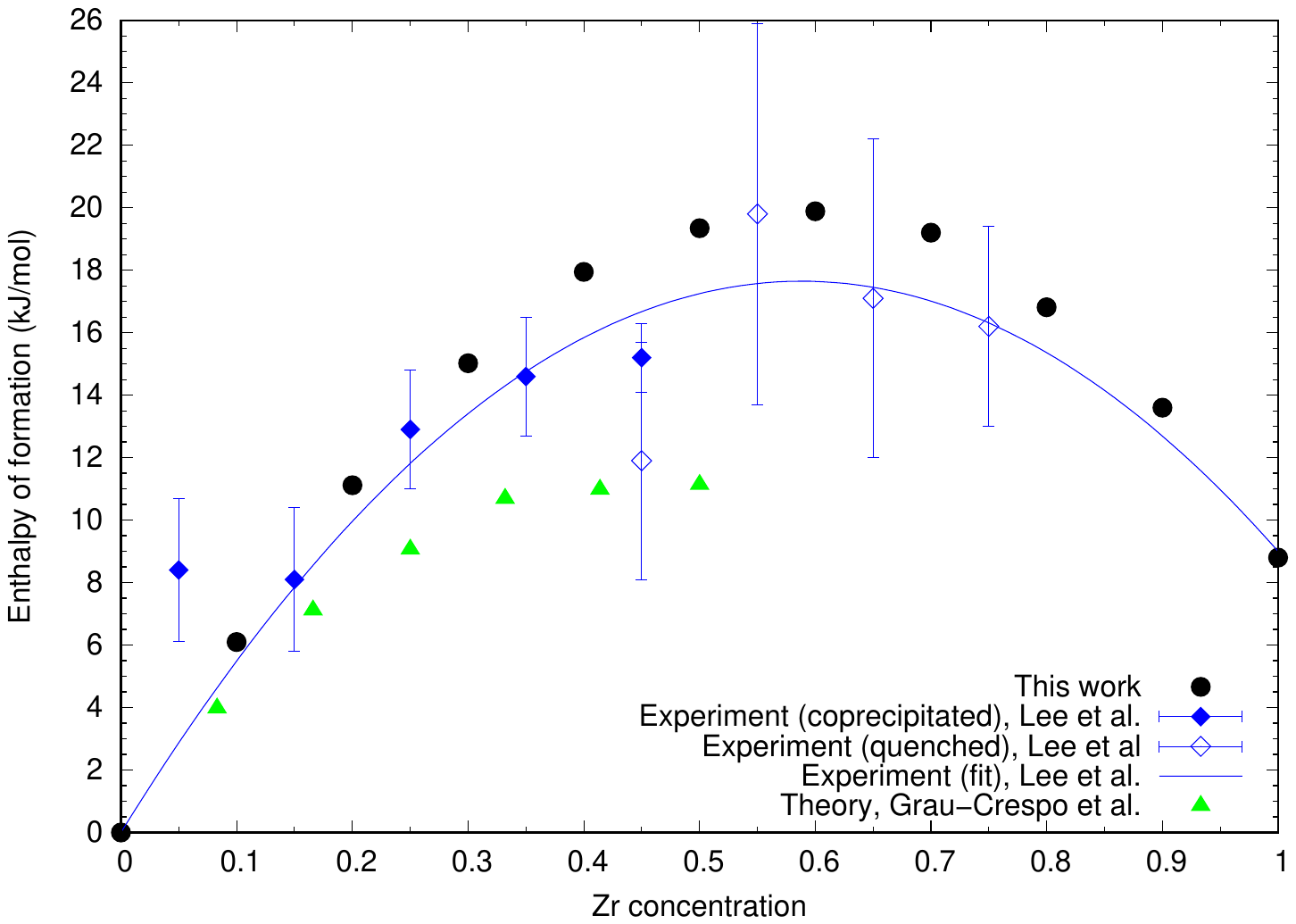}
	\caption{Enthalpy of mixing (Eqn.~\ref{eqn:H_form}) of fluorite ceria-zirconia solid solutions obtained from this work compared against
     previous experimental
	(Lee et al. \cite{Lee2008}) and theoretical (Grau-Crespo et. al \cite{Grau-Crespo2011}) studies. The uncertainty in the results for this work are smaller than
	the size of the data points. Experimental data from Lee et al. are given for different sample preparation methods (quenched, coprecipitated), as well as a 
	fitted curve to data, details of which can be found in Ref.~\onlinecite{Lee2008}.} 
	\label{fig:formation_enthalpy}
\end{figure}

The results of these calculations are shown in Figs.~\ref{fig:lattice_parameter} and \ref{fig:formation_enthalpy}. 
Fig.~\ref{fig:lattice_parameter} shows the lattice parameter versus $c$. It can be seen from this figure that the lattice parameter 
decreases with Zr concentration. This is due to the fact that Zr has a smaller atomic radius than Ce. It can also be seen that our model 
gives reasonable agreement with experiment. 
Fig.~\ref{fig:formation_enthalpy} shows the enthalpy of mixing versus $c$. Again, it can be seen that our model gives reasonable agreement with 
experiment. Interestingly, the theoretical results of Ref.~\onlinecite{Grau-Crespo2011} overestimate the lattice parameter and underestimate the 
formation enthalpies. Reasons for this have been discussed in that study.

\begin{figure}
	\includegraphics[width=\columnwidth]{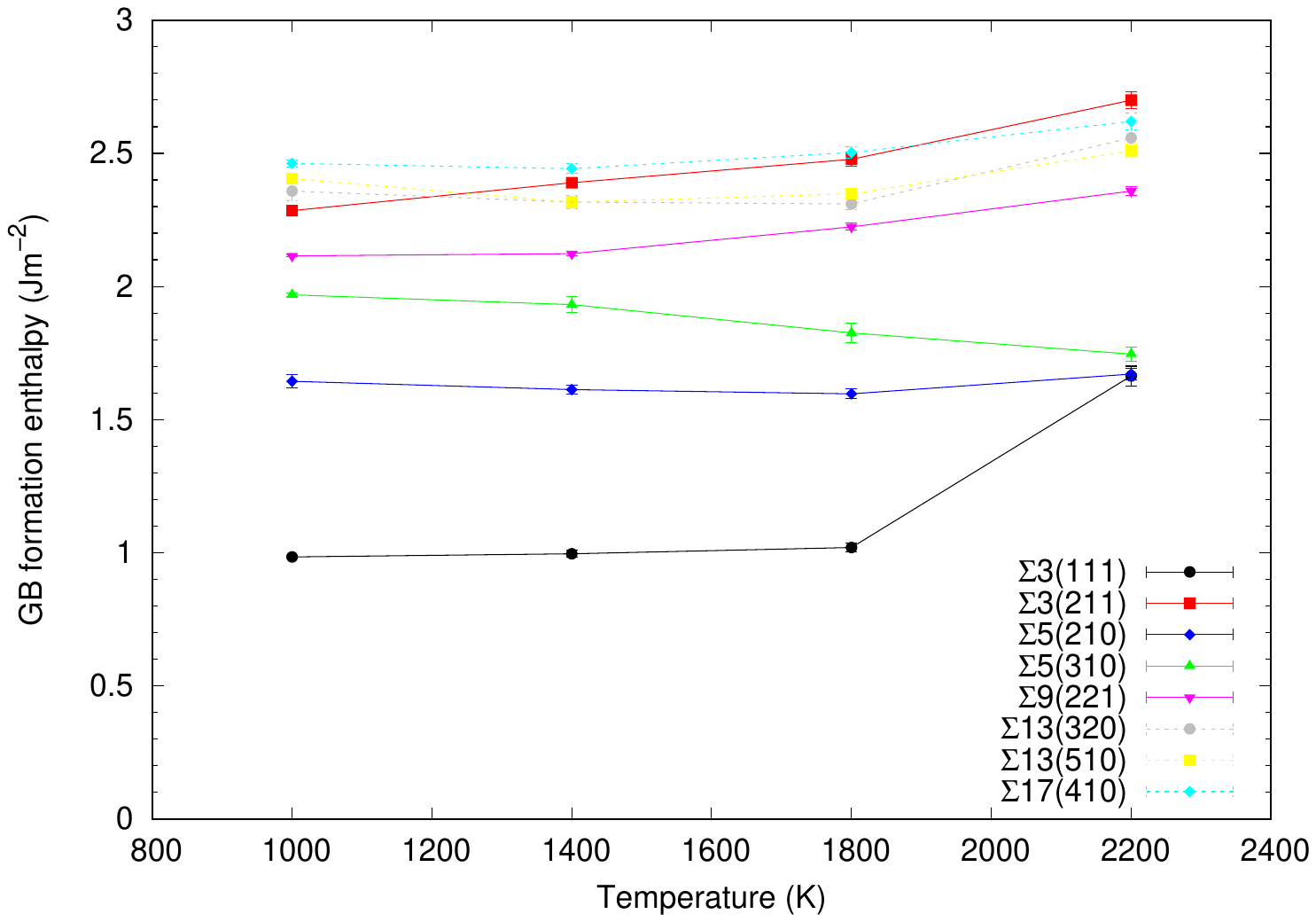}
	\caption{GB formation enthalpies for various GBs in pure fluorite ceria, at various temperatures. } 
	\label{fig:pure_GB_formation_enthalpies}
\end{figure}

\subsection{Critical point determination and Frenkel defects}
We initially considered the temperature range 1000K to 2500K for our bulk free energy calculations, hoping to pinpoint the critical (eutectoid) point. 
However, above 2200K our free energy calculations failed to converge. Upon further investigation, we found that this
was due to the spontaneous creation of Frenkel defects in the oxygen sublattice at these temperatures, defects which affected the ergodicity of the
simulations. For this reason in the main text we only present results for the range 1000K--2200K.

\begin{table}
    \begin{ruledtabular}
        \begin{tabular}{cccccc}
             $T$ (K) & $\Delta\mu_{\text{co}}$ (eV) & $c_1$ & $c_2$ \\
           \hline
1000 & -4.665(5) & 0.0018(1) & 0.99928(4) \\
1100 & -4.646(4) & 0.0033(1) & 0.99858(6) \\
1200 & -4.626(4) & 0.0062(2) & 0.9973(1) \\
1300 & -4.607(3) & 0.0101(3) & 0.9955(1) \\
1400 & -4.587(3) & 0.0153(4) & 0.9930(2) \\
1500 & -4.567(3) & 0.0228(6) & 0.9895(3) \\
1600 & -4.547(8) & 0.032(2)  & 0.985(1) \\
1700 & -4.527(3) & 0.045(1)  & 0.9787(6) \\
1800 & -4.506(4) & 0.061(2)  & 0.971(1) \\
1900 & -4.485(5) & 0.082(4)  & 0.962(2) \\
2000 & -4.463(2) & 0.109(2)  & 0.9493(8) \\
2100 & -4.441(4) & 0.145(6)  & 0.934(3) \\
2200 & -4.417(7) & 0.19(2)   & 0.916(6) \\
        \end{tabular}
    \end{ruledtabular}
    \caption{Properties of the miscibility gap for bulk ceria-zirconia obtained from free energy MC calculations and histogram reweighting,
    as described in the main text. $T$ is temperature; $\Delta\mu_{\text{co}}$ is the chemical potential difference corresponding to
    phase coexistence; and $c_1$ and $c_2$ are the binodal Zr concentrations of the two phases which coexist in the miscibility gap.
    The uncertainties quoted here are the size of change in $\Delta\mu$, $c_1$ and $c_2$ upon rewieighting from $\Delta\mu_{\text{sim}}$ to 
    $\Delta\mu_{\text{co}}$.} 
    \label{tab:coexistence_data}
\end{table}

\begin{figure*}
	\includegraphics[width=0.9\textwidth]{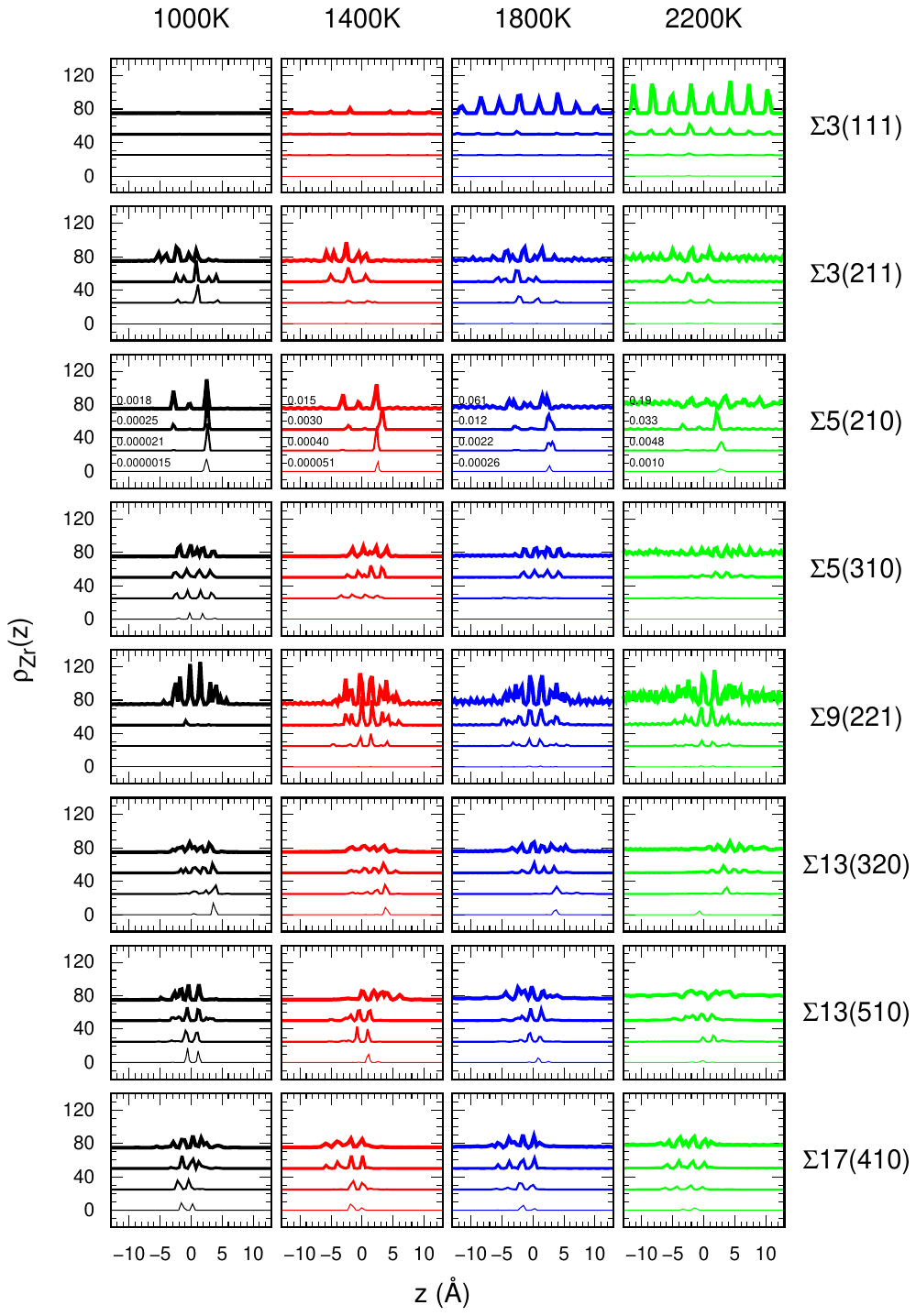}
	\caption{Zr density, $\rho_{\text{Zr}}(z)$ (Eqn.~\ref{eqn:seg_density}), versus distance perpendicular to the GB plane, $z$, for the
	GBs we considered at selected bulk compositions and temperatures. Each panel here corresponds to a particular GB and temperature: rows
    correspond to GBs and columns to temperatures, as labelled. Within each panel there are four curves, each of which corresponds to a
    different Zr concentration. In the row corresponding to the $\Sigma 5(210)$ GB the concentrations corresponding to all curves are labelled.
    Note that the same labelling scheme applies to all other rows. Note also that the curves corresponding to different concentrations are offset
    from one another, such that moving up the page corresponds to increasing the Zr concentration. $z=0$ corresponds to the GB plane.}
	\label{fig:zdensity}
\end{figure*}

\end{document}